\documentclass[12pt]{article}
\usepackage{amsmath,amssymb,amsfonts,graphicx}
\usepackage{slashed,tabularx}
\usepackage{young}
\usepackage{mathrsfs}
\usepackage[all]{xy}

\usepackage[usenames,dvipsnames]{xcolor}
\usepackage[colorlinks=true, pdfstartview=FitV, linkcolor=black, citecolor=black, urlcolor=black]{hyperref}

\def\leigh{Robert G. Leigh}
\def\weiss{Alexander B. Weiss}
\def\onkar{Onkar Parrikar}
\def\uiucaddress{\small Department of Physics, University of Illinois, 1110 W. Green St., 
Urbana IL 61801-3080, U.S.A. }
\def\title{\Large {The Exact Renormalization Group and Higher-spin Holography}}

\parskip=\baselineskip

\newcommand\cc[1]{#1^{^{\kern-6pt \circ}}\kern2pt}

\font\mybb=msbm10 at 11pt
\def\bb#1{\hbox{\mybb#1}}

\def\bR {\bb{R}}

\newcommand{\pa}{\partial}

\newcommand{\beq}{\begin{equation}}
\newcommand{\eeq}{\end{equation}}
\newcommand{\beqn}{\begin{eqnarray}}
\newcommand{\eeqn}{\end{eqnarray}}

\def\dalemb#1#2{{\vbox{\hrule height .#2pt
\hbox{\vrule width.#2pt height#1pt \kern#1pt
\vrule width.#2pt}
\hrule height.#2pt}}}

\newcommand{\cut}{M}
\newcommand{\rgla}{\cut d_{\cut}}
\newcommand{\bl}{{\boldsymbol\cdot}}
\newcommand{\Conn}{W}
\newcommand{\Scal}{B}
\newcommand{\ScalM}{\Pi}

\newcommand{\cScal}{\mathfrak{B}}
\newcommand{\cConn}{\mathcal{W}}
\newcommand{\cO}{\mathcal O}

\newcommand{\cL}{\mathcal{L}}
\newcommand{\re}{\mathbb{R}}

\newcommand{\Cont}{\widehat{W}}
\newcommand{\cCont}{\widehat{\mathcal{W}}}

\newcommand{\cD}{\mathcal{D}}

\newcommand{\alp}{\Delta}

\newcommand{\bd}{\boldsymbol d}
\newcommand{\cB}{\mathcal{B}}

\begin{document}

\begin{center}
\title
\end{center}
\vskip 1 cm
\centerline{
	{\bf
	{\leigh, \onkar\ and \weiss}}
	}
\vspace{.5cm}
\centerline{\it \uiucaddress}

\begin{abstract}
In this paper, we revisit scalar field theories in $d$ space-time dimensions possessing $U(N)$ global symmetry. Following our recent work\cite{Leigh:2014tza}, we consider the generating function of correlation functions of all $U(N)$-invariant, single-trace operators at the free fixed point. The exact renormalization group equations are cast as Hamilton equations of radial evolution in a model space-time of one higher dimension, in this case $AdS_{d+1}$. The geometry associated with the RG equations is seen to emerge naturally out of the infinite jet bundle corresponding to the field theory, and suggests their interpretation as higher-spin equations of motion. While the higher-spin equations we obtain are remarkably simple,  they are non-local in an essential way. Nevertheless, solving these bulk equations of motion in terms of a boundary source, we derive the on-shell action and demonstrate that it correctly encodes all of the correlation functions of the field theory, written as `Witten diagrams'. Since the model space-time has the isometries of the fixed point, it is possible to construct new higher spin theories defined in terms of geometric structures over other model space-times. We illustrate this by explicitly constructing the higher spin RG equations corresponding to the $z=2$ non-relativistic free field theory in $D$ spatial dimensions. In this case, the model space-time is the Schr\"odinger space-time, $Schr_{D+3}$. 
\end{abstract}

\pagebreak

%
\section{Introduction}

It is widely believed that gauge/gravity duality (or holography) should be interpreted as a geometrization of the renormalization group (RG) of quantum field theories. In this picture, scale transformations in the field theory correspond to movement in the extra `radial' direction, and specific RG trajectories correspond to specific geometries, which are asymptotically $AdS$ if the RG flow begins or ends near a fixed point. Early papers \cite{deBoer:1999xf,deBoer:2000cz} on the subject noted the relationship between RG flow and Hamilton-Jacobi theory of the bulk radial evolution. Additional contributions were made for example by \cite{Alvarez:1998wr,Akhmedov:1998vf,Schmidhuber:1999rb,Henningson:1998gx,Balasubramanian:1999re,Skenderis:2002wp} and more recent discussions include \cite{Koch:2010cy,Faulkner:2010jy,Heemskerk:2010hk,Lee:2009ij,Lee:2013dln,Gomes:2013qza}. 

From the perspective of quantum field theory, considerations of the renormalization group usually begin within the context of perturbation theory, naturally interpreted in terms of deformations away from a free RG fixed point. Indeed, the `exact renormalization group' (ERG) originally formulated by Polchinski  \cite{Polchinski:1983gv} was constructed within the confines of a path integral over bare elementary fields with (regulated) canonical kinetic terms corresponding to the free fixed point. Both the power and the curse of ERG is that it is formulated in terms of the free fixed point. One of the hallmarks of holography is that it pertains to a quite opposite limit, in which simple geometric constructions in the bulk correspond to strongly coupled dynamics in the dual field theory. So on the face of it, one might expect very little relationship to exist between the exact renormalization group and holography.

However, there exists a conjectured duality \cite{Klebanov:2002ja,Sezgin:2002rt,Leigh:2003gk} between free vector models in $d=2+1$ and certain types of higher-spin theories on $AdS_{4}$ (for a detailed exposition to higher-spin theories, see for e.g., \cite{Vasiliev:1995dn, Vasiliev:2012vf, Vasiliev:1999ba} and the reviews \cite{Bekaert:2005vh, Giombi:2012ms}). While the field theory side in this case is completely under control, the bulk is a far more complicated, and highly non-linear theory involving fields of arbitrarily high spin. Nevertheless, one might hope that this model provides an accessible testing ground for the holography/RG correspondence. A useful way to think of these vector model/higher-spin dualities can be illustrated by considering 3d Chern-Simons theories, known to be `dual' to 2d Wess-Zumino-Witten models. In this case, the theory is topological in the bulk (thus giving rise to a sort of holography long appreciated by condensed matter theorists (and experimentalists!)). What this means is that the theory does not depend on a bulk metric, and diffeomorphism invariance is broken only on the boundary through the introduction of boundary terms that explicitly involve a boundary metric. In particular, it is conjectured that 3d gravity\cite{Witten:1988hc,Witten:2007kt} (and higher spin generalizations\cite{Gaberdiel:2010ar,Gaberdiel:2010pz,Gaberdiel:2011zw}) can be thought of in these terms. Here, the dynamical degrees of freedom do not include a metric in the bulk, but at least a wide class of classical solutions have a geometric interpretation, the Chern-Simons gauge fields recast in terms of a co-frame and spin connection (or higher spin generalizations thereof). The equations of motion are first order (i.e., classical solutions are flat connections). We emphasize that one can interpret Chern-Simons theory in radial phase space terms, the precise details determined by the choice of a boundary action.  Given any such choice, different components of the connection correspond to `coordinates $q$' and `momenta $p$' of this phase space. Through the holographic dictionary, we expect $p$ and $q$ to correspond to expectation values and sources, respectively, for operators in the dual field theory. If we rewrite the bulk action in terms of $p,q$, the radial Hamiltonian is pure (Gauss) constraint, and the first order Chern-Simons equations of motion are nothing but Hamilton's equations. It is of course this latter structure that generalizes to other dimensions. As we will now review, the picture that emerges from a study of exact RG equations of vector models is that we should formulate their holographic duals in terms of connections and sections of certain bundles over a model space-time ($AdS$, for example). 

Indeed, following the initial proposal of \cite{Douglas:2010rc}, we considered in \cite{Leigh:2014tza} the Wilson-Polchinski exact renormalization group equations for a specific theory containing $N$ Majorana fermion fields in $2+1$ dimensions. More precisely, we considered the partition function of the theory as a function of sources for all \emph{bi-local}, single-trace, $O(N)$-invariant operators. Taking all such operators is a convenient way of organizing the infinite set of $O(N)$-invariant conformal modules with higher-spin quasi-primaries. Central to our construction was the recognized role of a huge symmetry (that we called $CO(L_2)$) of free field theories under which the elementary fields transform linearly but bi-locally. Since in a path integral formulation the elementary fields are not operators but just integration variables, changing integration variables by such a transformation relates the partition function evaluated at different values of the sources, leading to Ward identities. It is crucial in this construction that the theory is properly regulated (the path integral exists) and that the sources are written in an appropriate fashion. The beauty of the Majorana model was that (because the free action contains only one derivative) the appropriate structure was more or less manifest, and the sources could be immediately understood in terms of a connection for the $CO(L_2)$ group as well as a section (of an associated endomorphism bundle). It is this $CO(L_2)$ that becomes the `gauge group' of the corresponding higher spin theory. The renormalization group equations, interpreted as the equations of motion in one-higher dimensional RG mapping space, then provide the equations of motion for these sources. Remarkably, the ``higher-spin equations'' so obtained, are ostensibly simpler than those of Vasiliev higher-spin theory. The price we pay is that our RG equations are non-local in an essential and unavoidable way. It seems then, that \emph{the holographic dual to free field theory is most straightforwardly formulated in terms of non-local (in spacetime) variables}.

In the present paper, we apply these methods to complex scalar field theories (in arbitrary space-time dimensions $d>2$) with $U(N)$ global symmetry. The understanding of the geometric structure appearing in the Majorana model can be carried over to this case -- the $CO(L_2)$ symmetry described above translates into $CU(L_2)$. As we will show, the bosonic theory is somewhat simpler than the Majorana model, as the holographic phase space can be formulated entirely in terms of a scalar source and its conjugate momentum. This comes about through an extra (i.e., independent of the $CU(L_2)$ symmetry) redefinition symmetry possessed by the bosonic model. The ERG analysis gives rise to a complete Hamilton-Jacobi structure. We solve the Hamilton equations in terms of a boundary source, evaluate the corresponding on-shell action, and thus recover all $n$-point correlation functions of $U(N)$-singlet operators. These correlation functions are written holographically: each correlation function corresponds to a bulk `Witten diagram'. It is a simple matter to show that these reduce to the known correlation functions of the free fixed point, and correspondingly, the bulk on-shell action can be re-summed to reproduce the field theory generating function in the log det form, confirming that the holographic interpretation has lost no information about the free fixed point. 



One of the central ideas of holography is that the conformal symmetries of a field theory fixed point are reflected in the isometries of the bulk background geometry. This of course need not be $AdS$ --- $AdS$ pertains when the conformal symmetry is relativistic. To demonstrate this, in the final section we construct the exact RG equations for the $z=2$ non-relativistic free fixed point using light-cone quantization methods, and thus develop a higher spin gauge theory defined not on $AdS$, but on the Schr\"odinger geometry, $Schr_{D+3}$. 


\section{Overview: The free relativistic $U(N)$ model}
In this section, we review the prescription in \cite{Leigh:2014tza}, adapted to the case of $N$ complex scalar fields in $d$ space-time dimensions. The fixed point action is given by
\beq
S^0_{Bos.} = -\int d^dx\;\phi^*_m(x)\Box_{(x)} \phi^m(x)
\eeq
where we have taken the space-time metric to be $g_{\mu\nu} = \eta_{\mu\nu}$ and $\Box = \eta^{\mu\nu}\partial_{\mu}\partial_{\nu}$. Following \cite{Polchinski:1983gv}, we will regulate the action by introducing a smooth cutoff function $K_F(s)$ which has the property that $K_F(s) \to 1$ for $s < 1$ and $K_F(s) \to 0$ for $s >1$. A central object in \cite{Leigh:2014tza} was the regulated derivative operator
\beq
P_{F;\mu}(x,y) = K_F^{-1}(-\Box_{(x)}/M^2)\partial^{(x)}_{\mu}\delta^{d}(x-y),
\eeq
and we will use this here to construct a regularized action for the bosonic theory.
We are interested in deforming the theory away from the free fixed point with generic ``single-trace'' operators\footnote{Restricting to single-trace operators of course is not very general, but it is a {\it consistent truncation} of the full set of ERG equations in which sources for all ``multi-trace'' operators are included. We will return to this more general system in a subsequent publication \cite{Interactions}.}  of the schematic form
\beq\label{AllTheOps}
\phi_m^*\phi^m,\;\;\phi^*_m\pa_{\mu}\phi^m,\;\; \phi^*_m\pa_{\mu}\pa_{\nu}\phi^m,\cdots
\eeq
with no prejudice towards the number of derivatives. In order to do so, it is most convenient to introduce two bi-local sources $\Scal(x,y)$ and $\Conn_{\mu}(x,y)$.\footnote{\label{qle}While we will work with arbitrary bi-local sources for the most part, one might like to organize one's thoughts in terms of a \emph{quasi-local} expansion for these in the form
\beqn
\Scal(x,y) &\sim& \sum_{s=0}^{\infty} B^{a_1\cdots a_s}(x)\;\pa^{(x)}_{a_1}\cdots \partial^{(x)}_{a_s}\delta^d(x-y)+\cdots\label{QLE1}\\
\Conn_{\mu}(x,y) &\sim& \sum_{s=0}^{\infty} {\Conn_{\mu}}^{a_1\cdots a_s}(x)\;\pa^{(x)}_{a_1}\cdots \partial^{(x)}_{a_s}\delta^d(x-y)+\cdots\label{QLE2}
\eeqn
where the elipsis indicate more non-local (i.e. non-quasilocal) terms. Putting these expressions into the action, we see that they amount to sourcing arbitrary single-trace operators with any number of derivatives; such operators can be organized into conformal modules, represented by lowest weight quasi-primary operators.} 
We now write the full action for the $U(N)$ model as
\beq \label{action}
S^{reg.}_{Bos.}=-\int_{x,u,y}\phi^*_m(x)\eta^{\mu\nu}D_{\mu}(x,u)D_{\nu}(u,y) \phi^m(y)+\int_{x,y}\phi^*_m(x)\Scal(x,y)\phi^m(y)
\eeq
where we have introduced the notation
\beq
D_{\mu}(x,y) = P_{F;\mu}(x,y)+\Conn_{\mu}(x,y)
\eeq
One can easily check that this action sources all possible single-trace operators. We have written the action in this precise form, because, as in \cite{Leigh:2014tza}, we will see shortly that $D_{\mu}$ is a covariant derivative for a background gauge symmetry. The bilocal sources $B$ and $\Conn_{\mu}$ are really operators acting on $L_2$ functions over spacetime, but the bilocal representation is merely convenient notation which allows us to think in terms of `matrices'. Given this matrix notation, we will often use the ``dot'' notation for integration 
\beq
(f\bl g)(x,y) = \int_u f(x,u)g(u,y)
\eeq

The sources $\Scal$ and $\Conn_{\mu}$ that we have introduced above couple, respectively, to the following bi-local operators
\beq
\hat{\Pi}(x,y) = \phi^*_m(y)\phi^m(x),\;\;\;\;\hat{\Pi}^{\mu}(x,y) = \int_u\Big(\phi^*_m(y)D^{\mu}(x,u)\phi^m(u)-D^{\mu}(y,u)\phi^*_m(u)\phi^m(x)\Big)
\eeq
Note that $\hat\Pi^{\mu}(x,y)$ can be interpreted as a bi-local current operator. There is a minor subtlety in defining $U(N)$ singlet bilocal operators -- since $\phi^m(x)$ is a section of a $U(N)$ vector bundle, the only natural contraction between $\phi^*_m(y)$ and $\phi^m(x)$ should involve a $U(N)$ Wilson line. For instance,
\beq
\hat\Pi(x,y) = \phi_m^*(y){\left(\mathscr{P}\;e^{\int_y^xA^{(0)}}\right)^m}_n\phi^n(x) \label{u(n)WL}
\eeq
where $A^{(0)}$ is a background $U(N)$ connection. By not including the Wilson lines explicitly, we are assuming that the $U(N)$ vector bundle is trivial -- this means that $A^{(0)}$ can be taken to be flat, and in particular we make the choice $A^{(0)}=0$. 
 
The generating function (or partition function) is obtained by performing the path integral
\beq\label{PI}
Z[M, U, \Scal,\Conn] = \left(\mathrm{det}(-P^2_{F})\right)^N\int \left[d\phi d\phi^*\right]e^{iU + iS^{reg.}_{Bos.}}
\eeq
The path integration in \eqref{PI} is over the set of all square integrable complex scalar functions over the space-time $\re^{1,d-1}$, where the measure is conventionally written formally as
\beq\label{measure}
\left[d\phi d\phi^*\right] = \prod_{m=1}^N\prod_{x\in \re^{1,d-1}} d\phi_m(x) d\phi_m^*(x) 
\eeq
In the above, we have also introduced a source $U$ for the identity operator to keep track of the overall normalization, and a determinant normalization factor out front to ensure that the path-integral is well-defined in the presence of the cutoff function.

\subsection{The $U(L_2)$ and $CU(L_2)$ symmetries}
Given the measure in equation \eqref{measure}, it is natural to ask what a general linear transformation in function space would do to the path integral. To that end, consider a general linear bi-local field redefinition
\beq
\phi(x) \mapsto \int_y \cL(x,y)\phi(y)
\eeq
where $\cL: L_2(\re^d) \to L_2(\re^d)$ is a \emph{unitary} map of square integrable functions, i.e.
\beq
\cL^{\dagger}\bl \cL(x,y) \equiv \int_u \cL^{*}(u,x)\cL(u,y) = \delta^d(x-y).
\eeq
We will refer to the group of such transformations as $U(L_2(\re^d))$, or simply $U(L_2)$ for short.\footnote{Here we are considering such transformations that commute with the $U(N)$. This is appropriate since we are sourcing $U(N)\subset O(2N)$ singlets, and so we have a specific complex structure on the space of elementary fields. The sources themselves are of course real-valued.} If we consider an infinitesimal version of the above transformation
\beq
\cL(x,y) \simeq \delta(x-y)+\epsilon(x,y)
\eeq
then the $U(L_2)$ condition implies
\beq
\epsilon^*(x,y)+\epsilon(y,x)=0
\eeq
For example, consider an $\epsilon$ of the form
\beq
\epsilon(x,y) = i\xi(x)\;\delta(x-y)+\xi^{\mu}(x)\;\partial_{\mu}^{(x)}\delta(x-y)+i\xi^{\mu\nu}(x)\;\pa_{\mu}^{(x)}\pa_{\nu}^{(x)}\delta(x-y)+\cdots
\eeq
where $\xi,\;\xi^{\mu},\;\xi^{\mu\nu}\cdots$ are all real. This satisfies the $U(L_2)$ condition provided $\pa_{\mu}\xi^{\mu}=0,\;\pa_{\mu}\xi^{\mu\nu}=0$ and so on. The first term above is an infinitesimal $U(1)$ gauge transformation, the second term is a volume-preserving diffeomorphism, while the rest are higher-derivative transformations. 

Formally, the measure \eqref{measure} is invariant under $U(L_2)$ transformations, i.e. the Jacobian is unity. Coming to the action \eqref{action}, we obtain
\beq
S^{reg.}_{Bos.}[\cL\bl\phi, \Scal,\Conn_{\mu}] = S^{reg.}_{Bos.}\left[\phi,\;\cL^{-1}\bl\Scal\bl\cL, \;\cL^{-1}\bl\Conn_{\mu}\bl\cL+\cL^{-1}\bl\left[P_{F;\mu},\cL\right]_{\bl}\right]
\eeq
Thus, we find that $\Conn_{\mu}$ acts like a \emph{background} gauge field for unitary bi-local field redefinitions, while $\Scal$ conjugates tensorially. In the infinitesimal case, the transformation properties of $\Scal$ and $\Conn$ can be written as
\beq
\delta \Scal = \left[\Scal,\epsilon\right]_{\bl},\;\;\;\;\delta\Conn_{\mu}=\left[D_{\mu},\epsilon\right]_{\bl}
\eeq
where we have defined the `$\bl$-bracket' $\left[f,g\right]_{\bl}=f\bl g-g\bl f$. Given the formal invariance of the path integral measure, we obtain the Ward identity
\beq
Z[M,U,\Scal,\Conn_{\mu}] = Z\left[M,U,\;\cL^{-1}\bl\Scal\bl\cL,\; \cL^{-1}\bl\Conn_{\mu}\bl\cL+\cL^{-1}\bl\left[P_{F;\mu},\cL\right]_{\bl}\right]
\eeq
Note that ordinarily we would write the partition function for the free-fixed point in the above notation as $Z[M,U,0,0]$. But the $U(L_2)$ symmetry we encountered above teaches us a vital lesson -- since $\Conn_{\mu}$ behaves like a background connection under $U(L_2)$, the configuration $\Conn_{\mu}=0$ is gauge equivalent to the pure-gauge configuration $\Conn_{\mu} = \cL^{-1}\bl \left[P_{F;\mu},\cL\right]_{\bl}$. Thus, for any flat connection $\Conn^{(0)}$ satisfying
\beq
d\Conn^{(0)} + \Conn^{(0)}\wedge \Conn^{(0)}=0 \label{flatness}
\eeq
with $d = dx^{\mu}P_{F;\mu}$, the partition function $Z[M,U,0,\Conn^{(0)}]$ describes the free-fixed point. For this reason, we will find it convenient to pull out a flat piece from the full source $\Conn$ and write it as
\beq\label{connsplit}
\Conn = \Conn^{(0)}+\Cont
\eeq
Indeed, it is $\Cont$ and $\Scal$ which represent arbitrary single-trace, \emph{tensorial} deformations away from the free-fixed point, and thus parametrize single-trace RG flows away from the free CFT. We will return to this point shortly.

The group $U(L_2)$ does not exhaust the background symmetries of the free bosonic $U(N)$ vector model. We can further enlarge this group, by considering transformations of the form
\beq
\cL^{\dagger}\bl \cL(x,y) \equiv \int_u\cL^{*}(u,x)\cL(u,y)= \Omega^2(x)\delta^d(x-y) \label{cul2}
\eeq
where $\Omega$ is an arbitrary real function. In what follows, we will mostly focus on the case of constant $\Omega$ -- the general case should be a straightforward extension and its significance\footnote{In particular, for diffeomorphism invariance in the dual theory.} should not be neglected. We will call this larger group $CU(L_2)$.

Firstly, the measure of the path integral is, in general, not invariant under these transformations (unless $\Omega =1$), but will in general pick up an overall normalization factor, which one might think of as an anomaly. This can be absorbed into the source for the identity operator, which we will indicate by
\beq 
U \to \widehat{U}
\eeq
As we will see shortly, $CU(L_2)$ transformations induce a Weyl rescaling of the background metric. In order to continue thinking of the field theory as possessing Minkowski metric, we introduce a conformal factor $z$ in the background metric: $\eta_{\mu\nu}\mapsto z^{-2}\eta_{\mu\nu}$, and redefine the sources by rescaling them: $\Scal_{old} = z^{d+2} \Scal_{new}$ and $\Conn_{old}=z^d\Conn_{new}$.\footnote{\label{scalenote}These scale factors are put in for the following reason: recall that the sources admit the quasi-local expansions \eqref{QLE1}, \eqref{QLE2}. These expressions will get modified upon the introduction of the conformal factor $z$ in the metric, as $\delta^{(d)}(x-y) \to z^d\delta^{(d)}(x-y)$. The scale factors introduced above precisely remove this additional $z$-dependence. The extra $z^{-2}$ in $B_{new}$ ensures that it transforms tensorially under $CU(L_2)$.} For simplicity, we will drop the subscript  \emph{new} from here on. With these changes, the action takes the form
\beq
S^{reg.}_{Bos.}[\phi,M,z,\Scal,\Conn]=-\frac{1}{z^{d-2}}\int_{x,u,y}\phi_m^*(x)D_{\mu}(x,u)D^{\mu}(u,y)\phi^m(y)+\frac{1}{z^{d-2}}\int_{x,y}\phi_m^*(x)\Scal(x,y)\phi^m(y)
\eeq
where 
\beq
D_{\mu}(x,y) = K_F^{-1}(-z^2\Box_{(x)}/M^2)\partial^{(x)}_{\mu}\delta^{d}(x-y)+\Conn_{\mu}(x,y)
\eeq
and by $\Box_{(x)}$ we mean the $\eta$-d'Alembertian. We note that the effective ``renormalization scale'' now appears to be $\mu=M/z$. Indeed, as we will see in the following section, the renormalization group flow will be parametrized by $z$, while $M$ is essentially an auxiliary parameter inside the cut-off function, which sets the length scale . We will take $z$ to lie within the range $z\in [\epsilon,\infty)$, with $z=\epsilon$ corresponding to the ultraviolet cutoff $\Lambda_{UV} = \frac{M}{\epsilon}$, and $z\to \infty$ corresponding to the infra-red.

Having made these changes, we find straightforwardly
\beq
S^{reg.}_{Bos.}\Big[\cL\bl\phi,M,z,\Scal,\Conn_{\mu}\Big] = S^{reg.}_{Bos.}\Big[\phi,\lambda^{-1}M,\lambda^{-1}z,\;\cL^{-1}\bl\Scal\bl\cL,\;\cL^{-1}\bl\Conn_{\mu}\bl\cL+\cL^{-1}\bl\left[P_{F;\mu},\cL\right]\Big]
\eeq
where $\cL$ is a $CU(L_2)$ element satisfying equation \eqref{cul2}, with $\Omega = \lambda^{\frac{d-2}{2}}$. Once again, we find that the 1-form $\Conn_{\mu}$ transforms like a gauge field, while the 0-form $\Scal$ conjugates tensorially. Note further, that the conformal factor $z$ rescales to $\lambda^{-1}z$, and so does $M$. Thus, we conclude that $CU(L_2)$ is a background symmetry of the action up to a conformal rescaling of the background metric and the cutoff. In terms of the quantum partition function, we have the Ward identity
\beq
Z[M,z,\Scal,\Conn_{\mu},U] =Z [\lambda^{-1}M,\lambda^{-1}z,\;\cL^{-1}\bl\Scal\bl\cL,\;\cL^{-1}\bl\Conn_{\mu}\bl\cL+\cL^{-1}\bl\left[P_{F;\mu},\cL\right],\widehat U]
\eeq
Since the $CU(L_2)$ transformations involve a rescaling of the background metric, we expect them to play an important role in the renormalization group analysis. Indeed, this will be the case.

However, we have one more background symmetry to discuss before we move on to the renormalization group. Recall that we have split the 1-form $\Conn_{\mu}$ into a flat connection and a tensorial piece -- see eq. (\ref{connsplit}).
With this separation, the action becomes
\beq
S^{reg.}_{Bos.}[\phi,M,z,\Scal,\Conn_{\mu}] = S_{0}+S_{1}
\eeq
\beq
S_0=-\frac{1}{z^{d-2}}\int_{x,y,u}\phi^*_m(x)\eta^{\mu\nu}D^{(0)}_{\mu}(x,u)D^{(0)}_{\nu}(u,y)\phi^m(y)
\eeq
\beq
S_{1}=\frac{1}{z^{d-2}}\int_{x,y}\phi^*_m(x)\Big(B(x,y)-\left\{\Cont^{\mu},D^{(0)}_{\mu}\right\}_{\bl}(x,y)-\Cont_{\mu}\bl\Cont^{\mu}(x,y)\Big)\phi^m(y)
\eeq
where $D^{(0)}_{\mu} = P_{F;\mu}+\Conn^{(0)}_{\mu}$. Since $\Cont_{\mu}$ is tensorial, it is possible to redefine $B$ to absorb the terms involving $\Cont$
\beq
\cB = B -\left\{\Cont^{\mu},D^{(0)}_{\mu}\right\}_{\bl} -\Cont_{\mu}\bl\Cont^{\mu}
\eeq
More formally, one can write an identity for the partition function
\beq
Z[M,z,\Scal,\Conn^{(0)}_{\mu},\Cont_{\mu}+\Lambda_{\mu}]= Z[M,z,\Scal-\left\{\Lambda^{\mu},D_{\mu}\right\}_{\bl}-\Lambda_{\mu}\bl\Lambda^{\mu},\Conn^{(0)}_{\mu},\Cont_{\mu}]
\eeq
Therefore, one can use the above freedom to set $\Cont_{\mu}=0$ in $S_{1}$; we will henceforth do so, and write the deformations away from the fixed point as
\beq
S_{1}=\frac{1}{z^{d-2}}\int_{x,y}\phi^*_m(x)\cB(x,y)\phi^m(y)
\eeq
Note that this was in fact the starting point of Ref. \cite{Douglas:2010rc}, but the geometrical structure has now been made manifest. In our discussion of the exact RG equations to follow, were we not to absorb $\Cont_{\mu}$, we would find that the exact RG equation cannot unambiguously be separated into independent equations for $B$ and $\Cont_{\mu}$. The phase space of the dual theory is coordinatized entirely by fields whose boundary values are $B(x,y)$ and $\Pi(x,y)$. 

\subsection{Infinite jet bundles} \label{Sec2.2}

We have seen above that the large symmetry of free field theory, which is best elucidated in the path integral formulation, has a naturally geometric flavor. In particular, $\Conn$ -- which sources a certain bi-local current operator in the field theory -- transforms like a `connection'. As explained in \cite{Leigh:2014tza}, a natural interpretation for $\Conn$ is that it is a connection on the \emph{infinite jet bundle} of the field theory. Said another way, the background $U(L_2)$ and $CU(L_2)$ symmetries of free field theory can be characterized as gauge transformations of its infinite jet bundle, and sourcing all possible single-trace operators is equivalent to picking a connection on (and a section of the endomorphism bundle of) the infinite jet bundle corresponding to the field theory. For completeness, we will end this section by briefly recalling a few details of this construction -- this discussion is not strictly required to read the rest of the paper, and some readers might want to skip ahead to section 3. For a somewhat more technical discussion, see \cite{Leigh:2014tza}. 

While it is true that the $U(L_2)$ and $CU(L_2)$ symmetries we have discussed resemble gauge symmetries, the main problem we must confront in order for such an interpretation to hold, is their non-local nature. The gauge transformations one usually encounters in physics are local -- consider for instance a $U(1)$ gauge transformation $\delta\phi(x) = i\alpha(x)\phi(x)$. In this case, $\phi$ is thought of as a section of a vector bundle associated to a principal $U(1)$ bundle, and the gauge transformation may be thought of as a vertical group action. On the other hand, a $U(L_2)$ transformation 
\beq
\delta\phi^m(x) = \int_y \epsilon(x,y)\phi^m(y) \label{jet1}
\eeq
depends on the value of $\phi^m$ not merely at one point, but over the entire common support of $\epsilon$ and $\phi^m$. In other words, the action in \eqref{jet1} depends on the value of the $\phi^m$ at a point, and its derivatives at that point. In order to interpret this as a gauge transformation then, there is a need to construct a vector bundle whose fibre at each point keeps track of $\phi^m$, and its derivatives. In mathematics, this construction is referred to as the infinite jet bundle. Loosely speaking, the infinite jet bundle is a vector bundle whose fibre at a point $p$ consists of all equivalence classes of functions (or more generally sections) which have the same derivatives at $p$. Schematically, an element $\Phi$ of the fibre at $p$ correspondent to the function $\phi$ looks like
\beq
\Phi^m[\phi](p) = \left(\phi^m(x),\frac{\pa\phi^m}{\pa x^{\mu}}(p),\frac{\pa^2\phi^m}{\pa x^{\mu}\pa x^{\nu}}(p),\cdots\right)
\eeq
and is called the \emph{jet} of $\phi$ at $p$. The space of all jets at a point constitutes the fibre of the infinite jet bundle at that point. Going back to equation \eqref{jet1}, we see the action of $\epsilon$ on $\phi^m$ can be represented in terms of a linear and local action on its jet $\Phi^m[\phi]$. This is why we can think of $U(L_2)$ transformations as gauge transformations acting on the infinite jet bundle, satisfying the $U(L_2)$ condition. Given this interpretation, the 1-form $\Conn_{\mu}$ is naturally identified as a connection 1-form over the infinite jet bundle, while the 0-form $\Scal$ can be thought of as a section of its endormorphism bundle. Indeed, this interpretation fits nicely with our intuition for quasi-local expansions for our bi-local sources\footnote{These quasi-local expansions should be regarded as schematic. More precisely, we should think of the bilocal fields as sourcing all possible quasi-primary operators and their descendants, and hence the expansion is in terms of conformal modules.}
\beq
W_{\mu}(x,y) \simeq \sum_{s=1}^{\infty} W_{\mu}^{a_1\cdots a_{s-1}}(x)\pa_{a_1}^{(x)}\cdots\pa_{a_{s-1}}^{(x)}\delta^d(x-y)+\cdots \label{qle1}
\eeq
\beq
B(x,y) \simeq \sum_{s=1}^{\infty}B^{a_1\cdots a_{s-1}}(x)\pa_{a_1}^{(x)}\cdots\pa_{a_{s-1}}^{(x)}\delta^d(x-y)+\cdots
\eeq
The above quasi-local expansions basically express the fact that both $W_{\mu}$ and $B$ are valued in the endomorphism bundle of the jet bundle.\footnote{In most physics literature, the connection is thought of as a 1-form valued in the Lie-algebra of the gauge group, $W = W_{\mu}^{\alpha} T^{\alpha}dx^{\mu}$. The quasi-local expansions should be thought of in the same spirit, with the differential operators $T^{(s)}\simeq \pa^s_{(x)}\delta^d(x-y)$ playing the role of the Lie-algebra elements.} Of course, the jet bundle language is powerful enough to accomodate more general, non-local terms in $W_{\mu}(x,y)$ and $B(x,y)$, and this is indicated by the ellipsis in the above expansions. Indeed, as we will see later, the renormalization group forces the sources to become non-local in the infrared. In this way, a purely field theoretic exercise of sourcing all possible single-trace operators turns out to provide a beautiful geometric framework.

Before we proceed, we would like to introduce the notion of a \emph{Wilson line}. We define the Wilson line $\mathscr{K}_{\gamma}(t;t_0) $ along the curve $\gamma^{\mu}(s):[t_0,t]\to \re^d$ from the point $x_0$ to $x$ as the path ordered exponential 
\beqn
\mathscr{K}_{\gamma}(t;t_0) &=& \mathscr{P}_{\bl}\exp\;\int_{t_0}^tds\;\dot{\gamma}^{\mu}(s)W_{\mu}(s)\nonumber\\
&=& \boldsymbol{1}+\int_{t_0}^tds\;\dot{\gamma}^{\mu}(s)W_{\mu}(s)+\frac{1}{2}\int_{t_0}^tds_1\int_{t_0}^{s_1}ds_2\;\dot{\gamma}^{\mu}(s_1)W_{\mu}(s_1)\bl\;\dot{\gamma}^{\nu}(s_2)W_{\nu}(s_2)\nonumber\\
&+&\frac{1}{2}\int_{t_0}^tds_2\int_{t_0}^{s_2}ds_1\;\dot{\gamma}^{\nu}(s_2)W_{\nu}(s_2)\bl\;\dot{\gamma}^{\mu}(s_1)W_{\mu}(s_1)+\cdots
\eeqn
where $W_{\mu}(s) = W_{\mu}(\gamma(s),y)$ is the connection at the point $\gamma(s)$. Note here that the ``free index'' $y$ inside $W$ is an artifact of our bi-local notation -- it signifies that at each point along the curve, $W_{\mu}$ is a bilocal kernel (see equation \eqref{qle1} above). Also, the path-ordered exponential above is a ``$\bl$''-exponential, in that all the products involved in defining it are ``$\bl$''-products. As usual, the Wilson line defined above satisfies 
\beq
\frac{d}{dt}\mathscr{K}_{\gamma}(t;t_0) =  \dot{\gamma}^{\mu}(t)W_{\mu}(t)\bl\mathscr{K}_{\gamma}(t;t_0)
\eeq
An important property of Wilson lines is that if the connection $W$ is flat, then $\mathscr{K}_{\gamma}$ is independent of $\gamma$, and depends only on the endpoints. Some of these properties of Wilson lines will become relevant when we discuss the field theory correlation functions from a holographic point of view.  

\section{The Renormalization group and Holography}
Next, let us construct the renormalization group flow for the free bosonic vector model, perturbed away from the fixed point by the bi-local source $\cB$. In order to do so, we follow the conventional two-step process of Wilsonian RG:

\textbf{Step 1}: Lower the ``cutoff'' $M \to \lambda M$ (for $\lambda <1$), by integrating out a shell of ``fast modes'' -- this changes the sources, and we will use the notation $U\to \widetilde{U}$, $\cB \to \widetilde{\cB}$ to denote this. The calculation can be efficiently carried out using Polchinski's exact RG formalism (see Appendix A for details).

\textbf{Step 2}: Perform a $CU(L_2)$ transformation $\phi \to \cL\bl \phi$ to bring $M$ back to its original value, but in the process changing $z\to \lambda^{-1}z$ -- thus, the RG flow is parametrized by $z$ in our description, and not $M$ ($M$ is an auxiliary parameter in the cut-off function). The $CU(L_2)$ transformation additionally acts on the sources, and as we will see below, leads to a covariantization of the RG equations.  

The above two-step process can be succinctly stated in the form of the following equality of partition functions:
\beqn
Z[M,z,\cB, W^{(0)},U] &=& Z[\lambda M,z,\widetilde{\cB}, W^{(0)},\widetilde{U}]\label{RG1} \\
&=& Z[M,\lambda^{-1}z,\cL^{-1}\bl\widetilde{\cB}\bl\cL, \cL^{-1}\bl W^{(0)}\bl\cL+\cL^{-1}\bl\left[P_{F},\cL\right]_{\bl},\widehat{\widetilde{U}}]\label{RG2}
\eeqn
We can parametrize the infinitesimal RG transformation by writing $\lambda =1 - \varepsilon$, and
\beq
\cL = \mathbf{1}+\varepsilon\;zW^{(0)}_z
\eeq
where we have suggestively denoted the infinitesimal piece of $\cL$ as $W^{(0)}_z$, to indicate that it should be thought of as the $z$-component of the connection. From this point of view, $W^{(0)}_z$ is merely a book-keeping device which keeps track of the gauge transformations along the RG flow. Equations \eqref{RG1} and \eqref{RG2} then give us
\beq
W_{\mu}^{(0)}(z+\varepsilon z) = W_{\mu}^{(0)}(z)+\varepsilon z\left[P_{F;\mu}+W^{(0)}_{\mu},W^{(0)}_{z}\right]_{\bl}+O(\varepsilon^2)\label{conn1}
\eeq
\beq
\cB(z+\varepsilon z)=\cB(z)-\varepsilon z\left[W_z,\cB\right]_{\bl}+\varepsilon z\beta^{(\cB)}+O(\varepsilon^2)\label{scal1}
\eeq
\beq\label{WeylU}
U(z+\varepsilon z) = U(z)-i\varepsilon zN\mathrm{Tr}\;\Delta_B\bl \cB
\eeq
where the \emph{tensorial} beta function $\beta^{(\cB)}$ is given by (see Appendix A for details)
\beq
\beta^{(\cB)} = \cB\bl\Delta_B\bl\cB \label{beta}.
\eeq
We have also defined 
\beq
\Delta_B = \frac{M}{z}\frac{d}{dM}\left({D_\mu^{(0)}}^{-2}\right).
\eeq
By continuing the renormalization group flow, we can extend $\cB$ and $\Conn^{(0)}$ into the entire RG mapping space $\re^3\times \re^+$, where the half-line $\re^+$ is parametrized by $z$. We will often refer to this space as the \emph{bulk}, for reasons which will become apparent soon. We will also henceforth refer to the extended fields as $\cScal$ and $\cConn^{(0)}$, to emphasize that they live in the bulk. Note that $\cConn^{(0)}$ is a one-form in the bulk; indeed, $\cConn^{(0)}$ ``grows a leg'' in the $z$-direction, with $\cConn^{(0)}_z$ keeping track of the gauge transformations along the RG flow, as discussed above.

Comparing the $\varepsilon$ terms on both sides of equations \eqref{conn1} and \eqref{scal1}, we find
\beq
\partial_z\cConn^{(0)}_{\mu}-P_{F;\mu}\cConn^{(0)}_z+\left[\cConn^{(0)}_{z}, \cConn^{(0)}_{\mu}\right]_{\bl}=0 \label{conn2}
\eeq
\beq
\partial_z\cScal+\left[\cConn^{(0)}_z,\cScal\right]_{\bl} = \beta^{(\cScal)} \label{scal2}
\eeq
Therefore, the renormalization group equations emerge as gauge-covariant equations in the bulk. Given that $\cConn^{(0)}_{\mu}$ is also flat in the transverse directions (by construction; see discussion around equations \eqref{flatness} and \eqref{connsplit}), the first of these equations can be promoted to
\beq
\boxed{
\mathcal{F}^{(0)} \equiv \bd\cConn^{(0)}+\cConn^{(0)}\wedge \cConn^{(0)}=0} \label{conn3}
\eeq
where $\bd = dx^{\mu}\left[\pa_{\mu},\cdot\right]+dz \partial_z$ is the bulk exterior derivative. Note that the transverse component of the bulk exterior derivative is now simply $\pa_{\mu}$, instead of the regulated derivative $P_{F;\mu}$. The role of $P_{F;\mu}$ was to regulate the path integral; having extracted the RG equations, we will abruptly replace it with an ordinary derivative everywhere, except inside $\Delta_B$. One reason behind this choice is that it ensures $\boldsymbol{d}^2=0$. But a better justification will emerge in section \ref{sec:cfwd}, where we will show that the resulting bulk on-shell action reproduces precisely all the correlation functions of the boundary field theory, written as Witten diagrams. Moving on, equation \eqref{scal2} can similarly be promoted to a full-fledged one-form equation in the bulk
\beq
\boxed{
\cD_{(0)}\cScal\equiv \bd\cScal +\left[\cConn^{(0)},\cScal\right]_{\bl} = \boldsymbol{\beta}^{(\cScal)}}\label{scal3}
\eeq
The $z$-component of the one-form $\boldsymbol{\beta}^{(\cScal)}= \beta^{(\cScal)}_{\mu}dx^{\mu}+\beta^{(\cScal)}dz$ is given by equation \eqref{beta}; the transverse components on the other hand get determined by the Bianchi identity\footnote{The Bianchi identity is derived by acting on equation \eqref{scal3} with $\cD_{(0)}$, and using the fact that $\cConn^{(0)}$ is flat.}
\beq
\cD_{(0)} \boldsymbol{\beta}^{(\cScal)}\equiv \bd\boldsymbol{\beta}^{(\cScal)} +\left[\cConn^{(0)},\boldsymbol{\beta}^{(\cScal)}\right]_{\bl}=0 \label{Bianchi}
\eeq
Thus, the renormalization group equations for single-trace perturbations away from the free fixed point organize themselves into covariant equations, with the beta function playing the role of ``curvature''. In the following section, we will argue that these can be naturally interpreted as equations of motion describing the holographic dual of free field theory. It might be surprising that the equations we have derived above are remarkably simple, as compared to the Vasiliev higher spin equations. We emphasize that the equations are exact and form a consistent closed system. We have been able to establish these simple equations precisely because we have not insisted on locality. This is an essential aspect of free field theories.

%
Notably, equation \eqref{conn3} implies that the $\cConn^{(0)}$ is a \emph{flat} connection in the \emph{bulk}. This is where $AdS$ comes into the picture -- in the given coordinates, a particular solution (any other solution is of course gauge equivalent to this one) to the flatness condition is given by
\beq
\cConn^{(0)}(z;x,y) = -\frac{dz}{z}D(x,y)+\frac{dx^{\mu}}{z}P_{\mu}(x,y)\label{AdSconn}
\eeq
\beqn
D(x,y) &=& \frac{1}{2}\left(x^{\mu}\pa_{\mu}^{(x)}-y^{\mu}\pa_{\mu}^{(y)}+2\Delta_{\phi}\right)\delta^d(x-y),\;\;\;P_{\mu}(x,y)=\pa_{\mu}^{(x)}\delta^d(x-y)
\eeqn
which is easily identified as the $AdS$ connection\footnote{In evaluating the curvature of this connection, one should regard $P(x,y)$ and $D(x,y)$ as ``generators'' of the gauge group, and as such the exterior derivative $\boldsymbol{d}$ does not act on them.}. More precisely, equation \eqref{AdSconn} provides a $\mathfrak{g}=\mathfrak{o}(2,d)$-valued one form, which is in fact the Maurer-Cartan form for $O(2,d)$. By picking out an $\mathfrak{h}=\mathfrak{o}(1,d)$ subalgebra inside $\mathfrak{g}$, one identifies the corresponding $\mathfrak{h}$-valued part of $\cConn^{(0)}$ as the $AdS$ spin connection, while the remaining $\mathfrak{g}/\mathfrak{h}$-valued piece is identified as the $AdS$ co-frame. The fact that the isometry group of $AdS_{d+1}$ is precisely the conformal group $O(2,d)$ of $\re^{1,d-1}$  is, in the above language, manifested in the fact that there exists a subalgebra isomorphic to $\mathfrak{o}(2,d)$ inside the set of all gauge transformations which \emph{preserve} $\cConn^{(0)}$. 

We can similarly write down the Callan-Symanzik equations for $\Pi(x,y)$ following the two step RG prescription outlined above. We find (see Appendix \ref{app1} for details)
\beq
\Pi(z+\varepsilon z;x,y) = \Pi(z;x,y)-z\varepsilon\left[\Conn^{(0)}_z,\Pi\right]_{\bl}+i\varepsilon z N\Delta_B+\varepsilon z\mathrm{Tr}\;\gamma(x,y;u,v)\bl\Pi(v,u) \label{mom1}
\eeq
where we have introduced the notation
\beq
\gamma(x,y;u,v) = -\frac{\delta\beta^{(\cB)}(u,v)}{\delta \cB(y,x)}
\eeq
Note that $\Pi(x,y)$ transforms tensorially under $CU(L_2)$; we may extend it to a bulk adjoint-valued zero form $\mathcal{P}(x,y)$. Comparing $\varepsilon$ terms on both sides of equation \eqref{mom1}, we arrive at
\beq
\cD^{(0)}_z\mathcal{P} \equiv \pa_z\mathcal{P}+\left[\cConn^{(0)}_z,\mathcal{P}\right]_{\bl} = iN\Delta_B+\mathrm{Tr}\;\gamma(x,y;u,v)\bl \mathcal{P}(v,u) \label{CS1}
\eeq
Finally, we state the infinitesimal version of the RG ``Ward identities'' \eqref{RG1} and \eqref{RG2} explicitly
\beq
-\frac{\pa}{\pa z}Z = \mathrm{Tr}\left\{\left(\left[\cB,\Conn^{(0)}_z\right]_{\bl}+\beta^{(\cB)}\right)\bl\frac{\delta}{\delta\cB}+\left[D^{(0)}_{\mu},\Conn^{(0)}_z\right]_{\bl}\bl\frac{\delta}{\delta\Conn_{\mu}^{(0)}}\right\}Z+N\mathrm{Tr}\;\left(\Delta_B\bl\cB\right)Z \label{rgWI}
\eeq
where by $\frac{\pa}{\pa z}Z$ we mean the partial derivative with respect to $z$ keeping all the sources fixed. As we will see in the next section, this identity can be interpreted as the \emph{Hamilton-Jacobi} equation ($z$ being the parameter for ``radial evolution''), and plays a very crucial role in making contact with holography. 

\subsection{Holography and Hamilton-Jacobi theory}

In the previous section, we have seen how the renormalization group organizes field theory data in the one-higher dimensional RG mapping space. In this setup, the sources and the corresponding vacuum expectation values for single-trace deformations away from the fixed point turn into fields living in the bulk, with their dynamics governed by renormalization group equations. However, in order to ascribe a holographic interpretation to this, we must go further and show that all the correlation functions of the field theory can be reproduced from the bulk theory. The first step towards this, of course, is to construct the bulk action. 

The defining property of holography is contained in the following equation
\beq
Z[z_*, W^{(0)}(z_*),\cB(z_*)] = e^{iS_{HJ}[W^{(0)}(z_*),\cB(z_*)]} \label{Lifshitz}
\eeq
where $S_{HJ}$ is the \emph{Hamilton-Jacobi} functional for the bulk theory; i.e., the bulk action evaluated on-shell, with the boundary conditions $\cScal(z_*) = \cB(z_*)$. Said another way, the generating functional of the CFT is a wavefunctional (defined  on a constant $z=z_*$ hypersurface) from the bulk point of view in radial quantization. Therefore, while we might not have access directly to the bulk action, the field theory gives us the Hamilton-Jacobi functional instead. As is well-known from Hamilton-Jacobi theory, the (connected) boundary expectation value 
\beq
\Pi = \frac{\delta S_{HJ}}{\delta\cB}
\eeq
can be thought of as the boundary value of the momentum conjugate to $\cScal$ in the bulk. Thus, we see a bulk phase space picture emerging, with $\cScal$ and $\mathcal{P}$ forming a canonical pair. The canonical 1-form (of which the symplectic 2-form is the exterior derivative) is given by
\beq
\theta =\mathrm{Tr}\;\mathcal{P}\bl \delta\cScal \label{symplectic}
\eeq 
The crucial observation is that the RG Ward identity \eqref{rgWI} takes the form of the Hamilton-Jacobi equation
\beq
\frac{\pa}{\pa z}S_{HJ} = -\mathcal{H}
\eeq
with the bulk Hamiltonian given by
\beq\label{bulkHam}
\mathcal{H} = \mathrm{Tr}\left\{\left(\left[\cScal,\cConn^{(0)}_z\right]_{\bl}+\boldsymbol{\beta}_z^{(\cScal)}\right)\bl\mathcal{P}+\left[\cD^{(0)}_{\mu},\cConn^{(0)}_z\right]_{\bl}\bl\mathcal{P}^{\mu}\right\}-iN\;\mathrm{Tr}\;\left(\Delta_B\bl\cScal\right).
\eeq
It is straightforward to check that the Hamilton equations of motion which follow from the above are precisely the RG equations \eqref{conn2}, \eqref{scal2} and the Callan-Symanzik equations \eqref{CS1}. In addition to the above ``dynamical'' terms in the Hamiltonian, we may also introduce constraint terms, which enforce the transverse components (i.e., the $dx^{\mu}$ components) of equations \eqref{conn3}, \eqref{scal3}
\beq
\mathcal{H}_{constr.}= \mathrm{Tr}\;\left\{\left(\cD^{(0)}_{\mu}\cScal-\boldsymbol{\beta}_{\mu}^{(\cScal)}\right)\bl\mathcal{Q}^{\mu}+\mathcal{F}^{(0)}_{\mu\nu}\bl\mathcal{Q}^{\mu\nu}\right\}
\eeq
where $\mathcal{Q}^{\mu}$ and $\mathcal{Q}^{\mu\nu}$ are Lagrange multipliers. Note that the Hamiltonian is linear in momenta, and as such there is no distinction between phase space and configuration space formalisms. Nevertheless, we may construct a ``phase space action'' (``$p\dot q-H$") given by 
\beq\label{fullaction}
I = \int_{\infty}^{\epsilon} dz\;\mathrm{Tr}\left\{\mathcal{P}^I\bl\left(\cD^{(0)}_I \cScal - \boldsymbol{\beta}_I^{(\cScal)}\right)+\mathcal{P}^{IJ}\bl\mathcal{F}^{(0)}_{IJ}+iN\;\Delta_B\bl\cScal\right\}
\eeq 
where we have collected together ${\cal P}, {\cal Q}^\mu$ into ${\cal P}^I$, etc.
It is worthwhile noting that the first variation of this action reproduces all the RG and Callan-Symanzik equations (in the gauge where the Lagrange multipliers are set to zero). More importantly, we will now show that this action reproduces all the correlation functions of the boundary theory, from a holographic perspective (see also \cite{Douglas:2010rc} for a related, but different approach).

Before continuing, it is perhaps instructive to point to one feature of the bulk Hamiltonian that the reader may not have anticipated, namely that it is not `pure constraint', as might have been expected for a gravitational theory. We will in fact see in the next section that the trailing term in (\ref{bulkHam}) plays a crucial role in the holographic correspondence.

\subsection{Correlation functions and Witten diagrams}\label{sec:cfwd}

In order to compute the field theory correlation functions, we follow the standard prescription, i.e., we compute the bulk action on-shell, and extract the boundary generating functional from it, as per equation \eqref{Lifshitz}, with $z_*=\epsilon$. Note that the first two terms in the bulk action vanish on-shell; the only non-trivial contribution comes from the last term
\beq
I_{o.s} = -iN\int_{\epsilon}^{\infty} dz\;\mathrm{Tr}\left(\Delta_B\bl\cScal\right)
\eeq
where the minus sign comes from flipping the limits of integration. We remark that this term may be traced back to the anomalous transformation (\ref{WeylU}).
Since the field $\cScal$ above is a solution to the bulk equation of motion $\cD^{(0)}\cScal = \boldsymbol{\beta}^{(\cScal)}$, what we should do is solve this equation (along with the Callan-Symanzik equation), and substitute back into the action. But before we do that, we need to set up boundary conditions. Since we have two equations at hand, we need two boundary conditions. In the present context, one boundary condition presents itself naturally -- we fix the value of $\cScal$ at the boundary $z=\epsilon$:
\beq
\cScal(\epsilon;x,y) = b^{(0)}(x,y) \label{BC1}
\eeq 
From the field theory point of view, $b^{(0)}$ clearly has the interpretation of fixing the source at the ultraviolet cutoff. For the other boundary condition, we fix $\mathcal{P}$ in the infra-red:
\beq
\lim_{z\to \infty}\;\mathcal{P}(z;x,y) = 0 \label{BC2}
\eeq
This condition is of course consistent with the Hamilton-Jacobi structure (and the canonical 1-form \eqref{symplectic}), and is akin to the interior boundary condition one encounters regularly in holography.\footnote{For the variational principle to be well defined, we must either fix $\cScal$ on the boundary, or set $\mathcal{P}=0$ on the boundary, and thus the boundary conditions we have chosen are consistent with the variational principle without any additional boundary terms. While this choice of boundary conditions is natural in the present case, there are other boundary conditions which also have physically interesting interpretations \cite{Interactions}.}  

The equations at hand are non-linear; it is convenient (and perhaps physically more instructive) to solve them iteratively. Consequently, we introduce a formal organizing parameter $\alpha$, writing
\beq
{\cScal}=\alpha{\cScal}_{(1)}+\alpha^2{\cScal}_{(2)}+\cdots
\eeq
\beq
\mathcal{P} = \mathcal{P}_{(0)}+\alpha\mathcal{P}_{(1)}+\alpha^2\mathcal{P}_{(2)}+\cdots
\eeq
and we will solve the equations of motion order by order in $\alpha$, later setting $\alpha$ to one.
Let us focus on the $\cScal$ equation first. In fact, it suffices to focus on the $z$-component of the equation of motion, as the remaining components are automatically enforced by the Bianchi identity \eqref{Bianchi}. Then, we have
\beqn
\left[ {\cal D}^{(0)}_z,{\cScal}_{(1)}\right]_{\bl}&=&0\label{bulk1}\\
\left[ {\cal D}^{(0)}_z,{\cScal}_{(2)}\right]_{\bl}&=&{\cScal}_{(1)}\bl\Delta_B\bl {\cScal}_{(1)}\label{bulk2}\\
\left[ {\cal D}^{(0)}_z,{\cScal}_{(3)}\right]_{\bl}&=&{\cScal}_{(2)}\bl\Delta_B\bl {\cScal}_{(1)}+{\cScal}_{(1)}\bl\Delta_B\bl {\cScal}_{(2)}\label{bulk3}\\
&\vdots&\nonumber
\eeqn
We immediately see that the system of equations can be solved sequentially, with the solution of one equation (and all before it) determining the right-hand side of the next. 
The first equation (\ref{bulk1}) is homogeneous and has the solution 
\beq
{\cScal}_{(1)}(z;x,y)=\int_{x',y'}K(z;x,x') b_{(0)}(x',y') K^{-1}(z;y',y) \label{zeroth}
\eeq
where we have defined the \emph{boundary-to-bulk Wilson line}
\beq
K(z)=\mathscr{P}_{\bl}\exp \Big(-\int_{\epsilon}^{z}dz'\; \cConn_z^{(0)}(z')\Big)
\eeq
satisfying the equation
\beq
\pa_zK(z)+\cConn_z^{(0)}(z)\bl K(z) = 0
\eeq
This Wilson line should be interpreted in the terms we described in Section \ref{Sec2.2} above. As usual, we will surreptitiously write equation \eqref{zeroth} as 
\beq
{\cScal}_{(1)}(z)=K(z)\bl b_{(0)}\bl K^{-1}(z) \label{b1}
\eeq
in favor of compact notation. What we have done above, is to recognize that conjugating by $K$ (i.e., pulling back from a bulk point to the boundary) effectively converts the covariant derivative in (\ref{bulk1}) to $\pa_z$. Since $W^{(0)}$ is flat by its equation of motion, $K$ is independent of the path connecting the endpoints. 
At this order, the on-shell action is simply
\beq
I^{(1)}_{o.s.}=-iN\int_{\epsilon}^{\infty} dz\; \mathrm{Tr}\ \Delta_B\bl {\cScal}_{(1)}=-iN\int_{\epsilon}^{\infty} dz\; \mathrm{Tr}\ \left(K^{-1}\bl\Delta_B\bl K\bl b_{(0)}\right)\label{onshell1}
\eeq
It is convenient at this point to define the \emph{Wilsonian Green function} for the boundary field theory
\beq
g(z;x,y) = \int_{\epsilon}^{z}dz'\;H(z';x,y)=\int_{\epsilon}^zdz'\;\left(K^{-1}\bl \Delta_B\bl K\right)(z';x,y)\label{WGreenfunction}
\eeq
where
\beqn
H(z)\equiv K^{-1}(z)\bl \Delta_B(z)\bl K(z)=
\pa_z g(z)\label{WGreenfunction2}
\eeqn
and furthermore we will denote
\beq
g_{(0)}(x,y) \equiv g(\infty;x,y),
\eeq
which is in fact closely related to the free elementary field propagator of the boundary theory. To see this, note from the result (\ref{onshell1}) (or equivalently by solving the Callan-Symanzik equation at the zeroth order $\mathcal{D}^{(0)}\mathcal{P}_{(0)}=iN\Delta_B$, subject to the boundary condition eq. \eqref{BC2}), that 
\beq
\mathcal{P}_{(0)}(\epsilon;x,y)\equiv \langle \phi^*_m(y)\phi^m(x)\rangle_{CFT}=\frac{\delta I_{o.s.}}{\delta b_{(0)}(y,x)}\Big|_{b_{(0)}=0} = -iN\int_\epsilon^\infty dz\ H(z;x,y)=-iN\ g_{(0)}(x,y)
\eeq
where the subscript $CFT$ means the correlation function at the free-fixed point. This result implies that $\Delta_B$, which we defined earlier in the paper, can also be 
written \[\Delta_B=-\left[ \cD_{z}^{(0)}, \cD_{\mu}^{(0)}{}^{-2}\right]_{\bl}\] Thus we find that the linear term in the on-shell action can be written entirely in terms of boundary quantities
\beq
I^{(1)}_{o.s.}=-iN\; \mathrm{Tr}\;g_{(0)}\bl b_{(0)}.\label{OSaction1}
\eeq
The above computation can be represented in terms of a Witten diagram as in Fig. 1.
\begin{figure}[!h]
\centering
\includegraphics[height=5cm]{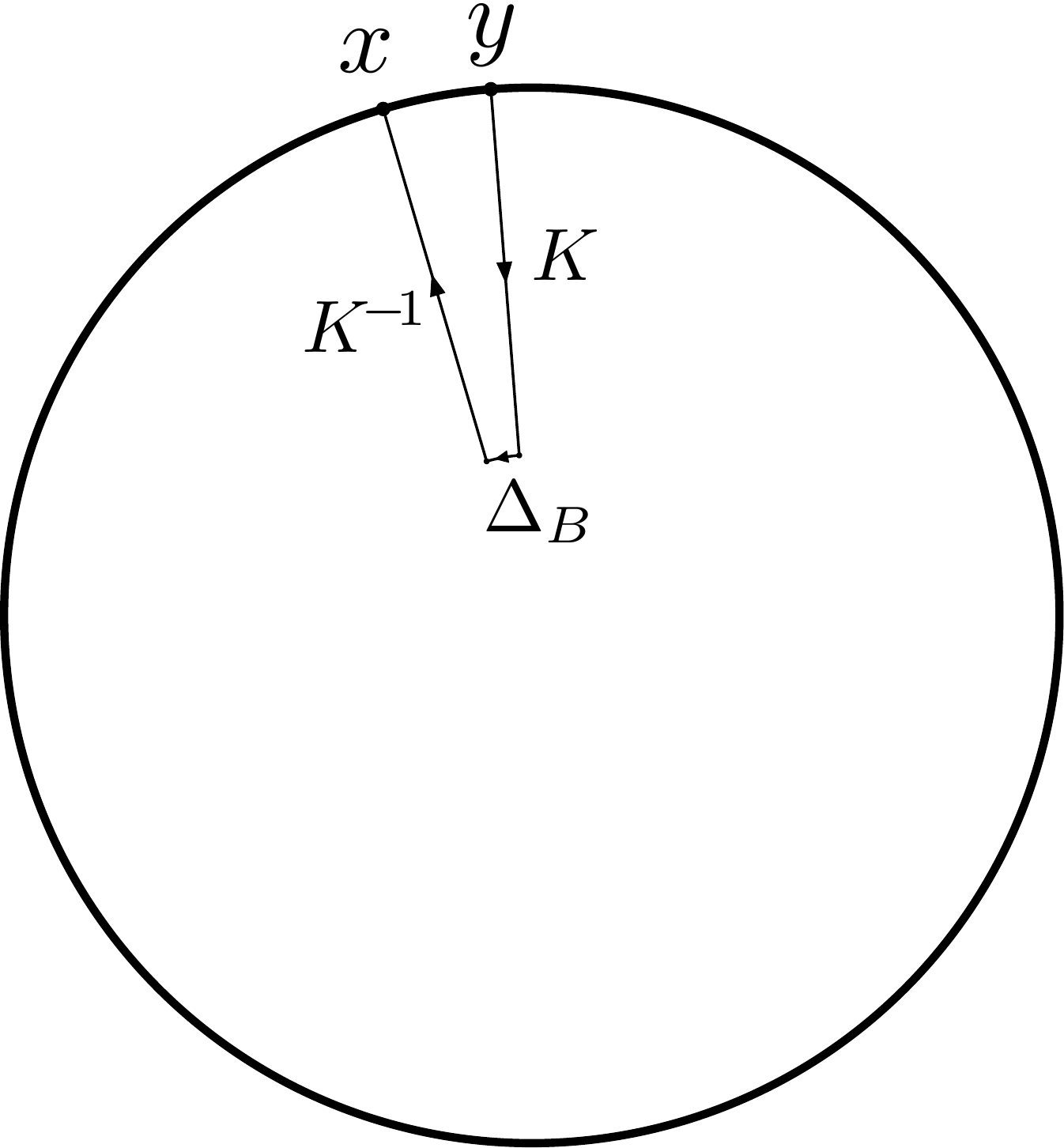}\label{WD1}
\caption{\small{The Witten diagram representation for the boundary one-point function $\mathcal{P}_{(0)}(x,y)$. The arrows indicate radial orientation, while the turn-around in the bulk represents an insertion of $\Delta_B$.}}
\end{figure}

Proceeding to second order, we solve equation \eqref{bulk2} with $\cScal_{(1)}$ given by equation \eqref{b1}
\beqn
\left[ {\cal D}^{(0)}_z,{\cScal}_{(2)}\right]_{\bl}= \Phi_{(2)}(z)&\equiv&K(z)\bl b_{(0)}\bl K^{-1}(z)\bl\Delta_B(z)\bl K(z)\bl b_{(0)}\bl K^{-1}(z)
\eeqn
More generally, at any given order, we can always write
\beqn
\left[ {\cal D}^{(0)}_z,{\cScal}_{(k)}\right]_{\bl}= \Phi_{(k)}(z)
\eeqn
where $\Phi_{(k)}$ is the inhomogenous term at the corresponding order. To solve this, we first conjugate by $K$ to reduce the covariant derivative to an ordinary derivative
\beq
K^{-1}(z)\bl\left[ {\cal D}^{(0)}_z,{\cScal}_{(k)}\right]_{\bl}(z)\bl K(z)=\pa_z\Big( K^{-1}(z)\bl {\cScal}_{(k)}(z)\bl K(z)\Big)
\eeq
and so we obtain
\beq
\pa_z\Big( K^{-1}(z)\bl {\cScal}_{(k)}(z)\bl K(z)\Big)=K^{-1}(z)\bl \Phi_{(k)}(z)\bl K(z)
\eeq
Taking without loss of generality the boundary condition to be $\cScal_{(k)}(\epsilon) = 0,\;\forall k\geq 2$ (since eq. \eqref{BC1} has been satisfied at first order in $\alpha$), the above equation can be easily solved
\beq
{\cScal}_{(k)}(z)=
K(z)\bl\left[\int_\epsilon^{\infty} dz'\ \Theta(z-z')\; K^{-1}(z')\bl \Phi_{(k)}(z')\bl K(z')\right]\bl K^{-1}(z)
\eeq
We can recognize here the \emph{ingoing bulk-to-bulk} Wilson line
\beq
G(z;z')=\Theta(z-z')\;K(z)\bl K^{-1}(z')=\Theta(z-z')\; \mathscr{P}_{\bl}\ \exp \Big(-\int_{z'}^{z}\;du\;\cConn_z^{(0)}(u)\Big)
\eeq
and the \emph{outgoing bulk-to-bulk} Wilson line
\beq
G^{-1}(z';z) = \Theta(z-z')\;K(z')\bl K^{-1}(z)=\Theta(z-z')\; \mathscr{P}_{\bl}\ \exp \Big(-\int_{z}^{z'}\;du\;\cConn_z^{(0)}(u)\Big)
\eeq
viz
\beq
{\cScal}_{(k)}(z)=
\int_\epsilon^{\infty} dz'\ G(z;z')\bl \Phi_{(k)}(z')\bl G^{-1}(z';z)
\eeq
Collecting everything together, we get the integral equation
\beq
{\cScal}(z)=K(z)\bl b_{(0)}\bl K^{-1}(z)+\int_\epsilon^{\infty} dz'\ G(z;z')\bl \beta^{(\cScal)}[{\cScal}](z')\bl G^{-1}(z';z)
\eeq
Returning to the second order calculation, we have
\beq
\cScal_{(2)} = \int_{\epsilon}^zdz'\;K(z)\bl b_{(0)}\bl K^{-1}(z')\bl \Delta_B(z')\bl K(z')\bl b_{(0)}\bl K^{-1}(z)
\eeq
and thus, the on-shell action at second order is given by
\beqn
I_{o.s}^{(2)} &=& -iN\int_{\epsilon}^{\infty}dz \int_{\epsilon}^{z}dz'\;\mathrm{Tr}\left(K^{-1}(z)\bl\Delta_B(z)\bl K(z)\bl b_{(0)}\bl K^{-1}(z')\bl \Delta_B(z')\bl K(z')\bl b_{(0)}\right)\\
&=& -iN\int_{\epsilon}^{\infty}dz \int_{\epsilon}^{z}dz'\;\mathrm{Tr}\left(H(z)\bl b_{(0)}\bl H(z')\bl b_{(0)}\right)
\eeqn
We can once again represent this in terms of a Witten diagram as in Figure 2. 
\begin{figure}[!h]
\centering
\includegraphics[height=5cm]{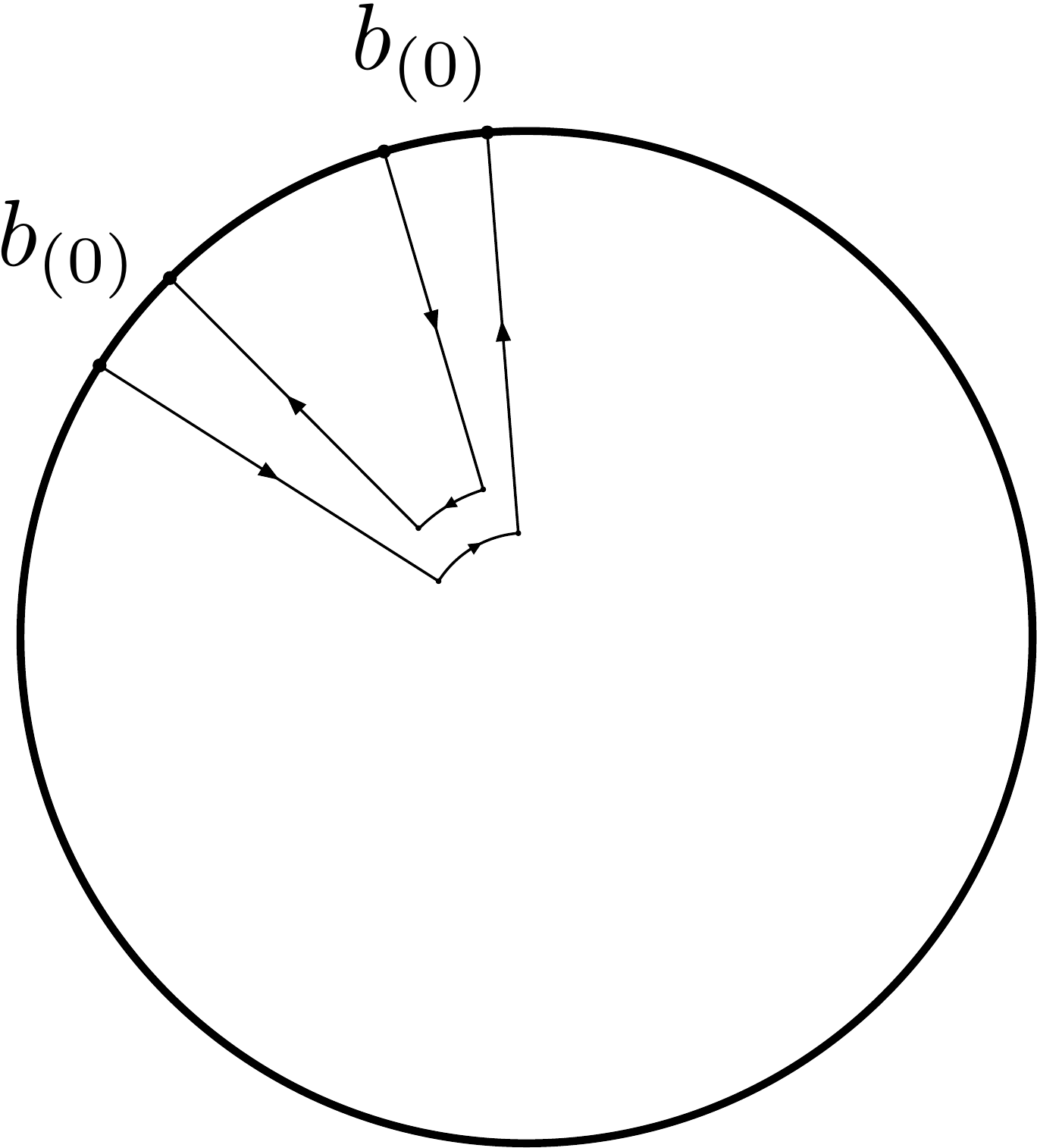} \label{WD2}
\caption{\small{The Witten diagram representing the second order term $I_{o.s}^{(2)}$ in the bulk on-shell action. The $b_{(0)}$s are boundary insertions of the ultraviolet bi-local source $b_{(0)}$. }}
\end{figure}
Using equation \eqref{WGreenfunction}, the $z$-integrations can be straightforwardly performed
\beqn
I_{o.s}^{(2)} 
&=&-iN\int_{\epsilon}^{\infty}dz \int_{\epsilon}^{z}dz'\;\mathrm{Tr}\left(H(z)\bl b_{(0)}\bl \pa_{z'} g(z')\bl b_{(0)}\right)\\
&=&-iN\int_{\epsilon}^{\infty}dz \mathrm{Tr}\left(\pa_z g(z)\bl b_{(0)}\bl g(z)\bl b_{(0)}\right)\\
&=&-i\frac{N}{2}\int_{\epsilon}^{\infty}dz\ \pa_z\mathrm{Tr}\left( g(z)\bl b_{(0)}\bl g(z)\bl b_{(0)}\right)
\eeqn
which integrates to
\beq
I_{o.s}^{(2)} = -i\frac{N}{2}\;\mathrm{Tr}\left(g_{(0)}\bl b_{(0)}\bl g_{(0)}\bl b_{(0)}\right).\label{OSaction2}
\eeq
This result reproduces the correct two-point functions of the free field theory. 

This procedure can be followed to arbitrary order. One finds the $k^{th}$-order term has the form
\beq
I^{(k)}_{o.s.}=-iN\int_{\epsilon}^{\infty}dz_1 \int_{\epsilon}^{z_1}dz_2...\int_{\epsilon}^{z_{k-1}}dz_k\;\mathrm{Tr}\left(H(z_1)\bl b_{(0)}\bl H(z_2)\bl b_{(0)}\bl...\bl H(z_k)\bl b_{(0)}+{\rm permutations}\right)
\eeq
The permutations include all of the distinct orderings of $\{H(z_2),...,H(z_k)\}$. Proceeding with the $z$-integrals as before, we find the on-shell action at this order is given by
\beq
I^{(k)}_{o.s.} = -i\frac{N}{k}\;\mathrm{Tr}\left(g_{(0)}\bl b_{(0)}\right)^k\label{OSaction3}
\eeq
As an example, the Witten diagram for the three point function is shown in Fig. 3. 
\begin{figure}[!h]
\centering
\includegraphics[height=5cm]{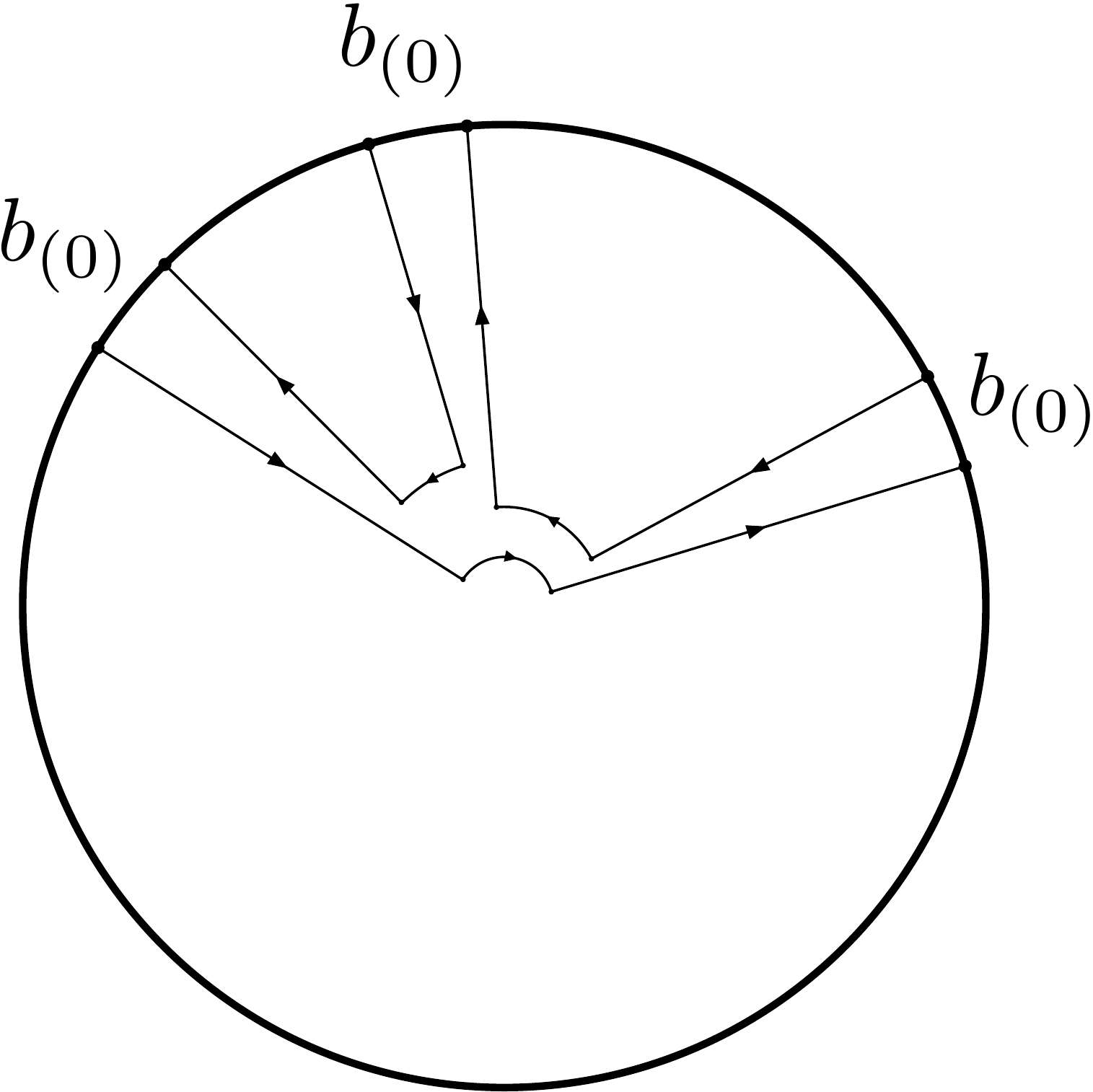}
\caption{\small{The Witten diagram for the bulk on-shell action at third order.}}
\end{figure}

Collecting equations \eqref{OSaction1}, \eqref{OSaction2}, \eqref{OSaction3}, we note that the on-shell action 
\beq
I_{o.s.} = -iN\Big(\mathrm{Tr}\left(g_{(0)}\bl b_{(0)}\right)+\frac{1}{2}\;\mathrm{Tr}\left(g_{(0)}\bl b_{(0)}\bl g_{(0)}\bl b_{(0)}\right)+\frac{1}{3}\;\mathrm{Tr}\left(g_{(0)}\bl b_{(0)}\bl g_{(0)}\bl b_{(0)}\bl g_{(0)}\bl b_{(0)}\right)+\cdots\Big)
\eeq
precisely reproduces the boundary generating functional 
\beq
Z[b_{(0)}]/Z[0]=e^{iI_{o.s.}}=\mathrm{det}^{-N}\left(1-g_{(0)}\bl b_{(0)}\right) \label{holpartitionfunction}
\eeq
Thus we conclude that the holographic formulation correctly reproduces all of the correlation functions of the boundary field theory. We will repeat the same analysis for the case of the fermionic vector model in Appendix \ref{app2}. 

Several comments are in order at this point. First, we have seen that a `double-line notation' naturally emerges for the Witten diagrams, essentially due to the bi-locality of the bulk field $\cScal$. However, because the connection $\cConn^{(0)}$ is flat, the corresponding Wilson lines can follow any path.\footnote{If the region between Wilson lines were filled in (as it would be in the presence of a dynamical $U(N)$ gauge field in the field theory) to obtain `open string worldsheets', the string tension would be zero.} Second, the `bulk vertex' is non-local. Each of these properties is a manifestation of unbroken higher spin symmetry at the free fixed point. Third, the above computation strengthens our claim that the action \eqref{fullaction} describes the holographic dual to the free bosonic vector model. It is only because the field theory in this case is completely under control, that we could construct the bulk holographic description by hand, and then check that we can go back and forth between the bulk and boundary descriptions. Finally, note from \eqref{holpartitionfunction} that our holographic description reproduced the ratio of partition functions $Z[b_{(0)}]/Z[0]$. $Z[0]$ is the domain of holographic renormalization. The divergences as $\epsilon\to 0$ contained in $Z[0]$ can be cancelled by local boundary counterterms.

\section{The Non-Relativistic $U(N)$ vector model}

The form of the higher spin theory that emerges from the exact renormalization group is in fact dictated by the assumed symmetries of the free fixed point. If we change those symmetries, we can expect to obtain a distinct higher spin holographic theory. We demonstrate here that this idea is correct by sketching the construction of the exact renormalization group equations for the free, non-relativistic $z=2$ bosonic theory (where $z$ is the Lifshitz scaling exponent). What one finds is that the analogue of the above discussion goes through, but the flat connection $W^{(0)}$ should be taken to be that of the Schr\"odinger geometry, as it is that connection which corresponds to a geometry with the isometries of the $z=2$ fixed point. (Related work in the context of Vasiliev higher spin theory may be found in Ref. \cite{Bekaert:2011cu}). The discussion in this section will not be as detailed as the previous sections, but is meant to sketch out the basic ideas involved. 

We have in fact almost all of the ingredients already assembled. The trick is to take the boundary d'Alembertian in the coordinatization
\beq
\Box=\pa_\xi\pa_t+\vec\nabla^2
\eeq
and assign the scaling symmetry $(\xi,t,\vec x)\mapsto (\xi,\lambda^2 t,\lambda x)$, which of course corresponds to dynamical exponent $z=2$. This is in fact the idea behind light-cone quantization (or DLCQ if the reader prefers) --- the non-relativistic theory is obtained from the relativistic theory in these specific coordinates. The program hangs together as long as the generator $N=\frac{\pa}{\pa\xi}$ is {\it central}, which allows for specifying a superselection sector of definite $N$ eigenvalue $n$ (and hence $\Box$ reduces to $in\pa_t+\vec\nabla^2$).\footnote{The $N$ eigenvalue would be identified with the non-relativistic mass in some particle interpretation (which here has no particular significance). The centrality of $N$ pertains for $z=2$ exclusively.}

We begin with a review of light-cone quantization in classical field theory. We take $N$ complex scalar fields $\phi^m$ with action
\beq
S=\int dtd^Dxd\xi\ \phi^\dagger_m \Box\phi^m
\eeq
where $\Box=\pa_\xi\pa_t+\vec\nabla^2$. Of course the theory is Lorentz invariant, but we write the d'Alembertian in these coordinates because we are going to do light-cone quantization, namely, we interpret $t$ as time. Notice that the model has a $U(N)$ spin-1 current with components
\beq
j_m{}^n=(\frac{i}{2}(\phi^\dagger_m\pa_\xi\phi^n-\pa_\xi\phi^\dagger_m \phi^n),i(\phi^\dagger_m\vec\nabla\phi^n-\vec\nabla\phi^\dagger_m \phi^n),\frac{i}{2}(\phi^\dagger_m\pa_t\phi^n-\pa_t\phi^\dagger_m \phi^n))
\eeq
The momentum conjugate to $\phi^m$ is $\pi_m=-\frac12\pa_\xi\phi^\dagger_m$ and the momentum conjugate to $\phi^\dagger_m$ is $\bar\pi^m=-\frac12\pa_\xi\phi^m$. We thus find a Hamiltonian of the form
\beq
H=\int d^Dxd\xi\ \vec\nabla\phi^\dagger_m\cdot \vec\nabla \phi^m
\eeq
and the charge operator (just the $U(1)$ charge for brevity) is 
\beq
Q=\frac{i}{2}\int d^Dxd\xi\ (\phi^\dagger_m\pa_\xi\phi^m-\pa_\xi\phi^\dagger_m \phi^m)
\eeq
The canonical equal-time commutation relations (ETCR) are
\beq
\left[ \phi^m(\vec x,\xi),\pa_\xi\phi^\dagger_{m'}(\vec y,\xi')\right]=-2i\delta^m_{m'} 2\pi\delta(\xi-\xi')(2\pi)^D\delta^{(D)}(\vec x-\vec y)
\eeq
It is this result that is the first indication that this theory has all the features of a non-relativistic field theory. We can mode expand the fields as
\beq
\phi^m(\vec x,\xi)=\int \frac{d^D\vec{p}}{(2\pi)^D}\int \frac{dn}{2\pi} e^{i\vec p\cdot\vec x}e^{in\xi} 
a^m_{n,\vec p}
\eeq
and to reproduce the ETCR, we have
\beq
\left[a^m_{n,\vec p},a_{m';n',\vec q}^\dagger\right] = -\frac{4\pi}{n}\delta^m_{m'}\delta(n-n')\ (2\pi)^D\delta^{(D)}(\vec p-\vec q)
\eeq
We would introduce a tensor Fock space ${\cal H}=\otimes_{n,\vec p} {\cal F}_{n,\vec p}$ and define lowest weight representations by 
\beq
 a^m_{n',\vec p'}|0\rangle\equiv a^m_{n',\vec p'}\otimes_{n,\vec p}|0\rangle_{n,\vec p}=0.
 \eeq
The excited states are then obtained by acting with $a^\dagger$'s.
Given this, we can identify the number operator
\beq
N_{m;n,\vec p}=-\frac{n}{2}a_{m;n,\vec p}^\dagger a^m_{n,\vec p}\quad {\rm (no\ sum)}
\eeq
and the charge operator is just
\beq
Q=\int \frac{d^D\vec{p}}{(2\pi)^D}\int \frac{dn}{2\pi}\ N_{m;n,\vec p}
\eeq
The Hamiltonian gives
\beq
Ha_{m;n,\vec p}^\dagger |0\rangle= \vec p^2a_{m;n,\vec p}^\dagger |0\rangle
\eeq
 We see that the quantum number $n$ represents a degeneracy -- the energies do not depend on it. The dynamics of the model do not distinguish $n$, and we can think about a given value of $n$ as corresponding to a superselection sector. 

Thus, at least in this free theory, we see that a superselection sector of the $d=D+2$-dimensional relativistic theory quantized on the lightcone is equivalent to a $D+1$-dimensional non-relativistic theory. In a given superselection sector labelled by $n$, we may then take the action to be
\beq
S_0 = \int dtd^{D}\vec{x}\;\phi_m^{\dagger}\left(in P_{F;t}+\vec{P}_F^2\right)\phi^m
\eeq
where now the regulated derivative is defined in a Schr\"odinger covariant manner
\beq
P_{F;\mu} = K_{F}^{-1}\left(-\frac{1}{M^2}(in\pa_t+\vec{\nabla}^2)\right)\pa_{\mu}
\eeq
This action represents a (regulated) free, non-relativistic conformal field theory \cite{Nishida:2007pj,Balasubramanian:2008dm}.  

However, let us be a little more specific about what we mean by a superselection sector. In terms of the naive free field theory, we mean that the action and path integral measure are sums and products respectively of quantities with fixed $n$,
\beq
Z=\prod_{n\in spec(N)} Z_n
\eeq
When we introduce the bi-local sources, we choose to preserve this structure. Given then a source term of the schematic form 
\beq
\int d\xi dtd^Dx\int d\xi' dt'd^Dx'\ \phi^\dagger_m(\xi,t,\vec x) B(\xi,t,\vec x;\xi',t',\vec x')\phi^m(\xi',t',\vec x')
\eeq
we see that requiring $\phi^m(\xi,t,\vec x)=e^{in\xi}\phi^m(n,t,\vec x)$ for fixed $n$ (i.e. requiring that the source does not mix different superselection sectors) is equivalent to requiring that within a superselection section labelled by $n$
\beq\label{srcxidep}
B(\xi,t,\vec x;\xi',t',\vec x')=e^{in(\xi-\xi')}B_n(t,\vec x;t',\vec x')
\eeq
Thus, had we taken the sources to be local in $\bR^{D+2}$, then they would have had no dependence on $n$. This is to be expected from non-relativistic holography: the bulk fields that give rise to conserved currents in the boundary field theory are gauge fields with $n=0$. We see though that in the higher spin version, this will be a little more subtle --- the higher spin fields will have some memory of $n$. 

In formulating the exact RG equations and thus the higher spin bulk theory, we will keep $\xi$-dependence and the $\xi$-components of the connection. However, ultimately what we will mean by ``DLCQ in the bulk" is that the bulk fields will be taken after the fact to have definite $\xi$-dependence, of the form (\ref{srcxidep}) (but see also the discussion below).

Keeping in mind our experience in dealing with the free relativistic models, we perturb away from this fixed point by introducing the bi-local sources $B(x,y), W_\mu(x,y)$:
\beq
S = \phi_m^{\dagger}\bl\left(D_{\xi}\bl D_{t}+\vec{D}^2\right)\bl\phi^m+\phi_m^{\dagger}\bl B\bl\phi^m
\eeq
where we have defined
\beq
D_{\mu}= P_{F;\mu}+W_\mu
\eeq
as before. This action of course transforms under $U(L_2(\re^{1,D+1}))$ transformations 
\beq
\phi_m\to \cL\bl\phi_m,\;\;\;\; \cL^{\dagger}\bl\cL=1
\eeq
with the sources transforming as 
\beq
B \to \cL^{-1}\bl B \bl \cL,\;\;\;\; W_{\mu}\to \cL^{-1}\bl W_{\mu}\bl \cL +\cL^{-1}\bl\left[P_{F;\mu},\cL\right]_{\bl} \label{UL2}
\eeq
From our discussion of DLCQ above, the next task is to construct $U(L_2(\re^{1,D+1}))$ transformations which do not mix various superselection sectors. This means that in any given sector labelled by $n$, we must restrict the $\xi$-dependence of our transformations as
\beq
\cL_n(\xi,t,\vec{x};\xi',t',\vec{x}') = e^{in(\xi-\xi')}\tilde{\cL}_n(t,\vec{x};t',\vec{x}')
\eeq
where $\tilde{\cL}_n$ satisfy
\beq
\tilde{\cL}_n^{\dagger}\bl \tilde{\cL}_n = \delta(t-t')\delta(\vec{x}-\vec{x}')
\eeq
We will refer to these subgroups as $U_n(L_2(\re^{1,D}))$.

Generalizing this to $CU(L_2(\re^{1,D+1}))$ is also straightforward, provided one is mindful of the fact that the coordinates are taken to scale as $(\xi,t,\vec x)\mapsto (\xi,\lambda^2 t,\lambda \vec x)$. As before, we introduce a conformal factor $z$ in the field theory metric
\beq
g^{(0)} = \frac{1}{z^2}\left(d\xi dt+d\vec{x}^2\right)
\eeq
Defining the dimensionless sources $W_{old} = z^{D+2}\;W_{new}$, and $B_{old}=z^{D+4}B_{new}$, the action becomes
\beq
S = \frac{1}{z^{D}}\;\phi_m^{\dagger}\bl\left(iD_{\xi}\bl D_{t}+\vec{D}^2\right)\bl\phi^m+\frac{1}{z^{D}}\phi_m^{\dagger}\bl B\bl\phi^m
\eeq
With these definitions, the action is invariant under $CU(L_2)$ transformations satisfying
\beq
\cL^{\dagger}\bl \cL(\xi,t,\vec{x};\xi',t',\vec{x}') =\lambda^{2\Delta_{\phi}}\delta(\xi-\xi')\delta(t-t')\delta^D(\vec{x}-\vec{x}')
\eeq
where $\Delta_{\phi} = \frac{1}{2}D$. The sources transform according to equation \eqref{UL2}, along with $z\to \lambda^{-1}z$. As in the relativistic case, we have the freedom to take $W_{\mu}=W^{(0)}_{\mu}$ to be flat. 

The renormalization group equations take the same form
\beq
\boldsymbol{d}\cConn^{(0)}+\cConn^{(0)}\wedge \cConn^{(0)}=0 \label{NRrg1}
\eeq
\beq
\boldsymbol{d}\cScal+\left[\cConn^{(0)},\cScal\right]_{\bl} = \boldsymbol{\beta}^{(\cScal)} \label{NRrg2}
\eeq
where $\boldsymbol{\beta}^{(\cScal)}=\cScal\bl\Delta_B\bl\cScal$, and $\Delta_{B;n}=\frac{M}{z}\frac{d}{dM}\left(iD^{(0)}_{\xi}\bl D^{(0)}_{t}+\vec{D}^{(0)2}\right)^{-1}$. 

Let us now project onto a superselection sector. Recall from our previous discussion, that in order to do this consistently, we must choose a specific $\xi$-dependence for the source, namely
\beq
\cScal(\xi,t,\vec{x}; \xi',t',\vec{x}') = e^{in(\xi-\xi')}\cScal_n(t,\vec{x}; t',\vec{x}')
\eeq
Note though that this $\xi$-dependence in $\cScal$ can be completely removed by a gauge transformation 
\beq
\cL_n(\xi,t,\vec{x};\xi',t',\vec{x}') = e^{in\xi}\delta(\xi-\xi')\delta(t-t')\delta^D(\vec{x}-\vec{x}')
\eeq
The renormalization group equations however transform covariantly under a gauge transformation, and so we arrive at the equations projected onto a superselection sector
\beq
\boldsymbol{d}\cConn_n^{(0)}+\cConn_n^{(0)}\wedge \cConn_n^{(0)}=0 \label{NRrg1}
\eeq
\beq
\boldsymbol{d}\cScal_n+\left[\cConn^{(0)}_n,\cScal_n\right]_{\bl} = \boldsymbol{\beta}^{(\cScal)}_n \label{NRrg2}
\eeq
where now $\boldsymbol{d}$ is the exterior derivative on $\re^{1,D}$, and $\cConn_n^{(0)} = (\cL_n^{-1}\bl\cConn^{(0)}\bl\cL_n+\cL_n^{-1}\bl\boldsymbol{d}\cL_n)$. The $z$-component of the $\beta$ function is given by
\beq
\boldsymbol{\beta}^{(\cScal)}_{n;z} = \cScal_n\bl\Delta_{B;n}\bl \cScal_n
\eeq
where now $\Delta_{B;n}=\frac{M}{z}\frac{d}{dM}\left(inD^{(0)}_{t}+\vec{D}^{(0)2}\right)^{-1}$. These are the final non-relativistic higher spin renormalization group equations. 

A natural choice for the flat connection is given by
\beq
\cConn_n^{(0)} = -\frac{dz}{z}D(x,y)+\frac{dt}{z^2}H(x,y)+\frac{d\vec{x}^i}{z}\vec{P}_i(x,y)+d\xi N(x,y)
\eeq
where we have used the shorthand notation $x=(t,\vec{x})$ and in the given superselection sector $N(x,y)$ evaluates to $in\delta^{(D+1)}(x,y)$. This last term above is important -- it ensures that the connection $\cConn^{(0)}_n$ is gauge equivalent to a connection of the form
\beq
e^{in(\xi-\xi')}\widetilde{\cConn}^{(0)}_n(x,y)
\eeq
which is crucial for a consistent projection onto superselection sectors. Note that the only remaining dependence on $n$ is in the $d\xi$ component of the connection. Indeed, the theory remembers the superselection sector through the holonomy of $\cConn^{(0)}$ around the $\xi$ cycle
\beq
\oint_{\xi-cycle} \cConn^{(0)}_n = in \delta^{(D+1)}(x-y)
\eeq
The remaining components of $\cConn^{(0)}_n$ give us the Schr\"odinger connection, which is the Maurer-Cartan form on the Schr\"odinger group. One can show, similar to the above analysis, that the Hamilton-Jacobi formalism can be constructed for this theory, and the Schr\"odinger-covariant correlation functions obtained.

\section{Discussion}

In this article, we have applied the techniques developed in \cite{Leigh:2014tza} to the case of free bosonic vector models. The essence of the construction was to interpret the covariant higher-spin exact renormalization group equations, written in terms of non-local variables, as the bulk equations of motion. We have then  computed the bulk on-shell action iteratively in terms of a general boundary source, and we showed that this reproduces precisely the field theory generating functional for connected correlation functions. Thus, taking advantage of the fact that free field theory is exactly solvable, we have been able to build a holographic dual, and demonstrate that we can go back and forth between the intrinsic field theory and the bulk holographic descriptions. What remains to be done is to to clarify the relationship between our higher spin theory and that of Vasiliev. On the face of it, the Vasiliev equations are written in terms of a large number of auxiliary variables, and are local in spacetime. The first steps towards a  map between the two formalisms were taken in \cite{Leigh:2014tza}, but a complete picture remains elusive. As a first step of course, it might be worthwhile reproducing Fronsdal equations from the higher-spin renormalization group equations. Given that our bilocal source $\cScal(x,y)$ sources all possible quasi-primary operators and their descendants, the coefficient functions obtained through its quasi-local expansion naturally organize in terms of conformal modules
\beq
\cScal(x,y) \in \sum_{s=0}^{\infty}D(2-s,s)
\eeq 
with the lowest weight states in each module being the Fronsdal fields. Indeed, since the Fronsdal wave operator is essentially the quadratic Casimir for the conformal group $SO(2,d)$ written in a bulk basis, it seems natural that the Fronsdal equations should follow as a natural consequence of conformal symmetry. The details of this will be presented elsewhere. 

While the above is interesting in its own right, a major motivation is indeed to be able to adopt some of these methods to the case of interacting field theories. Within the context of vector models, it has long been conjectured that the critical Bosonic/Fermionic $U(N)$ models (namely the non-trivial interacting IR/UV fixed points) are dual to type A/B Vasiliev theories with a different quantization for the bulk scalar.\footnote{In phase space terms, the `quantizations' are expected to be related by a canonical transformation.} In the $N\to \infty$ limit, the generating functionals of these interacting theories can be obtained as a Legendre transform with respect to the source of the scalar operator $\phi^*_m\phi^m$ (or $\bar{\psi}_m\psi^m$ for the fermionic case) -- thus, in principle, one can extract the holographic Hamiltonian, and hence the bulk equations of motion for the interacting case straightforwardly from our results, through a Legendre transform. However, it would be much more satisfying to apply the exact renormalization group  to the interacting fixed point, and demonstrate from from first principles the emergence of (massless) higher spin equations at large $N$. At finite $N$, one expects to see that all spin $s>2$ gauge fields are higgsed, leaving behind only the $s\leq 2$ fields. We will present the details of this investigation elsewhere\cite{Interactions}.

Another possible direction for future work is to generalize our ideas to the larger arena of theories with gauge symmetries, for example the Chern-Simons-vector models. The basic idea is that the singlet sector of free vector models discussed in this paper can be thought of as the $k\to \infty$ limit of a corresponding Chern-Simons vector model. Interestingly, there are conjectured holographic dualities between certain parity-breaking Vasiliev theories and finite-$k$ Chern-Simons vector models \cite{Giombi:2011kc,Chang:2012kt}.  
 
Finally, we note that an interesting connection between exact renormalization group and string field equations was sought to be made in \cite{Bank1987733}. Indeed, it is widely speculated that the higher-spin gauge fields are some sort of an effective description of a certain subsector of the higher-spin modes associated with string theory, which become massless in the $\alpha'\to \infty$ limit. Given the similarities between the ideas presented here and in \cite{Bank1987733}, it will be interesting to see whether or not this connection to string theory can be made more explicit.

\section{Acknowledgments}

We would like to thank Djordje Minic, Tassos Petkou, Leopoldo Pando Zayas, Diana Vaman, Matthias Gabardiel, Jaume Gomis, Rajesh Gopakumar, Kewang Jin, Cheng Peng, Ergin Sezgin and Joe Polchinski for interesting discussions. Some support for this research was provided by the U.S. Department of Energy, contract DE-FG02-13ER42001.

\appendix\numberwithin{equation}{section}
\section{Renormalization group: Details}\label{app1}
In this appendix, we spell out the details of the derivation of the exact renormalization group equations. We take the regulated action to be
\beq
S_{Bos.} = S_{0}+S_{1}
\eeq
\beq
S_{0} = -\frac{1}{z^{d-2}}\int_{x,y,u}\phi^*_m(x)D^{(0)}_{\mu}(x,y)D^{(0)}_{\mu}(y,u)\phi^m(u)
\eeq
\beq
S_{1}= \frac{1}{z^{d-2}}\int_{x,y}\phi^*_m(x)\cB(x,y)\phi^m(y)+U
\eeq

Next, we run the two-step RG process:

\textbf{Step 1}: In Step 1 of RG, we want to integrate out a shell of fast modes, and investigate how that changes the sources. In order to perform this integration, we use Polchinski's exact RG formalism. We start by lowering $M\to \lambda M$, where $\lambda = 1-\varepsilon$. Since this has the interpretation of integrating out fast modes, we can extract the change in the sources $\delta_{\varepsilon}\cB=-\varepsilon M\frac{d}{dM}\cB$ and $\delta U =-\varepsilon M\frac{d}{dM}U$ by imposing
\beq
M\frac{d}{dM} Z=Z_0^{-1}\int[d\phi\; d\phi^*]\left\{\left(M\frac{d}{dM}e^{iS_0}\right)e^{iS_{1}}+e^{iS_0}\left(M\frac{d}{dM}e^{iS_{1}}\right)-Z_0^{-1}e^{iS_0+iS_1}M\frac{d}{dM}Z_0\right\}=0\label{dMZ}
\eeq
where the last term above is from the normalization of the partition function\footnote{We have defined $Z_0 = \int[d\phi\; d\phi^*]e^{iS_0}$.}, as in (\ref{PI}). Evaluating the first term, we find
\beqn
M\frac{d}{dM}e^{iS_0} &=& -\frac{i}{z^{d-2}}e^{iS_0}\int\phi_m^*\bl\left(M\frac{d}{dM}D_{(0)}^2\right)\bl\phi^m\nonumber\\
&=& \frac{i}{z^{d-2}}e^{iS_0}\int \phi^*_m\bl D^{(0)}{}^2\bl \Delta_{B}\bl D^{(0)}{}^2\bl\phi^m\nonumber\\
&=&-iz^{d-2}\int_{x,y}\Delta_{B}(x,y)\left\{\frac{\delta^2}{\delta\phi^m(x)\delta\phi_m^*(y)}-i\frac{\delta^2S_0}{\delta^m\phi(x)\delta_m\phi^*(y)}\right\}e^{iS_0}. \label{dMS0}
\eeqn 
where we have defined $\Delta_{B} = M\frac{d}{dM}(D^{(0)}_{\mu}D^{(0)}_{\mu})^{-1}$. The second term in \eqref{dMS0} cancels with the contribution from the normalization. Therefore, integrating by parts from equations \eqref{dMZ} and \eqref{dMS0}, we are left with 
\beq
M\frac{d}{dM}e^{iS_{1}}-iz^{d-2}\int_{x,y}\Delta_{B}(x,y)\frac{\delta^2}{\delta\phi^*_m(x)\delta\phi^m(y)}e^{iS_{1}}=0
\eeq
Evaluating this term by term, we find 
\beq
i\left\langle M\frac{d}{dM}U+\frac{1}{z^{d-2}}\phi_m^*\bl M\frac{d}{dM} \cB\bl\phi^m\right\rangle=-\left\langle N\mathrm{Tr}\;\Delta_{B}\bl \cB+\frac{i}{z^{d-2}}\phi_m^*\bl \cB\bl \Delta_{B}\bl \cB\bl \phi^m\right\rangle
\eeq
As the notation suggests, the above equations should be regarded as valid inside the path integral. From the above equation, we can now read off the change in the sources (if we treat the 1 and $\phi_m^*(x)\phi^m(y)$ as independent operators)
\beq
\delta_{\varepsilon}\cB = -\varepsilon M\frac{d}{dM} \cB = \varepsilon\;\cB\bl\Delta_B\bl \cB
\eeq
\beq
\delta_{\varepsilon} U = -\varepsilon M\frac{d}{dM} U = -i\varepsilon N\;\mathrm{Tr}\;\Delta_B\bl \cB
\eeq

\textbf{Step 2}: Next in step 2, we perform a $CU(L_2)$ transformation
\beq
\label{RGstep2COL2}
\cL(x,y) = \delta^d(x-y)+\varepsilon\;zW^{(0)}_z(x,y)
\eeq
to bring the cutoff back while changing the conformal factor of the metric. Having done this, we label the sources $\cB(z), \cB(z+\varepsilon z)$ and $U(z), U(z+\varepsilon z)$. Together with step 1, we thus conclude
\beq
\cB(z+\varepsilon z)=  \cB(z)-\varepsilon\left[W^{(0)}_z, \cB\right]_{\bl}+\varepsilon\; \cB\bl\Delta_B\bl \cB
\eeq
\beq
U(z+\varepsilon z)=U(z)-i\varepsilon N\;\mathrm{Tr}\;\Delta_B\bl \cB
\eeq
In this way, the renormalization group extends the sources defined at a given value of $z$ to all of the bulk RG mapping space. Redefining $\Delta_B$ as $\Delta_B = \frac{M}{z}\frac{d}{dM}\left(D_{(0)}^2\right)^{-1}$, we recover equations \eqref{scal1} and \eqref{WeylU}. 

\subsection*{Callan-Symanzik equations}
Similarly, we can derive an expression for the Callan-Symanzik equations of the bi-local operator $\hat\ScalM(x,y) = \phi^*_m(y)\phi^m(x)$.  Again we run the two step RG process:

\textbf{Step 1}:  Defining normalized correlation functions by
\beq
	\langle \cO \rangle \equiv \frac{ \int [d\phi \; d\phi^*] \, \cO \, e^{i S} }{ \int [d\phi \; d\phi^*] \, e^{i S} }
\eeq
it is straightforward to demonstrate the relationship
\beq
	\rgla \ScalM
	\equiv \rgla \left\langle \hat\ScalM \right\rangle
	= \mathrm{Tr} \left\{ \alp_B \bl
	\left\langle
	 \frac{ \delta S_1 }{ \delta \phi_m^* } \frac{ \delta \hat\ScalM }{ \delta \phi^m }
	+ \frac{ \delta \hat\ScalM }{ \delta \phi_m^* } \frac{ \delta S_1 }{ \delta \phi^m }
	- i \frac{ \delta^2 \hat\ScalM }{\delta \phi_m^* \delta \phi^m}
	\right\rangle
	\right\}
\eeq
The right hand side can be calculated explicitely.  The result is
\beq
	\delta_{\varepsilon}\Pi=-\varepsilon\rgla \ScalM
	= i\varepsilon\; N \,z \alp_B
	-\varepsilon z\; \alp_B \bl \cB \bl \ScalM
	-\varepsilon z\; \ScalM \bl \cB \bl \alp_B
\eeq
or more compactly,
\beq
	\delta_{\varepsilon}\ScalM
	= i \varepsilon z\; N \, \alp_B
	+ \varepsilon z\;\mathrm{Tr} \left\{ \gamma \bl \ScalM  \right\}
\eeq
where we define
\beq
	\gamma(x,y;u,v)
	\equiv -\frac{ \delta \beta^{(\cB)} (u,v) }{ \delta \cB (y,x) }
	= -\delta(x-u) \left( \alp_B \bl \cB \vphantom\sum \right) \! (y,v)
	- \left( \cB \bl \alp_B \vphantom\sum \right) \! (u,x) \, \delta(v-y)
\eeq

\textbf{Step 2}:  We perform a $CU(L_2)$ transformation as given in \eqref{RGstep2COL2}.  The result is
\beq
	\ScalM(z+\varepsilon z;x,y)
	= \ScalM(z;x,y) - \varepsilon \, z \left[ \Conn^{(0)}_z , \ScalM \right]_{\bl} + i \varepsilon z\; N \, \alp_B+\varepsilon \, z \, \mathrm{Tr} \left\{ \gamma(x,y;u,v) \bl \ScalM(v,u) \right\}
\eeq
As with the beta function derived above, this relationship can be extended into the bulk.  Denoting the bulk momentum as $\mathcal{P}$, we have 
\beq
	\cD^{(0)}_z \mathcal{P}
	\equiv \pa_z \mathcal{P} + \left[ \cConn^{(0)}_z , \mathcal{P} \right]_{\bl}
	= i N \, \alp_B+\mathrm{Tr} \left\{ \boldsymbol\gamma(x,y;u,v) \bl \mathcal{P}(v,u) \right\}
\eeq
where $\boldsymbol\gamma(z;x,y;u,v) \equiv -\frac{ \delta \boldsymbol{\beta}^{(\cScal)} (z;u,v) }{ \delta \cScal (z;y,x) }$ is the bulk extension of $\gamma$.

\section{The Dirac fermion: Correlation functions}\label{app2}

In this section, we present some details of the $U(N)$ fermionic vector model at its free fixed point. The exact renormalization group for the $O(N)$ Majorana free fermion and its interpretation as a higher-spin holographic system was discussed in detail in \cite{Leigh:2014tza}. Here we present a  review of the Dirac fermion, with emphasis on reproducing the boundary correlation functions from the holographic action. 

The regulated Dirac action, with bi-local sources for $U(N)$-singlet, single-trace operators is given by
\beq
S_{Dirac} = \int_{x,y}\;\bar\psi(x)i\gamma^{\mu}P_{F;\mu}(x,y)\psi(y)+ \int_{x,y}\bar\psi(x)\left(A(x,y)+\gamma^{\mu}W_{\mu}(x,y)\right)\psi(y)
\eeq
Unlike the bosonic version, the structure of sources for the fermionic vector model is dimension dependent -- for instance, in $d=2n+1$ dimensions, one has the following single trace bi-local operators
\beq
\bar\psi_m(x)\psi^m(y),\;\;\bar\psi_m(x)\gamma^{\mu_1}\psi^m(y), \;\;\bar\psi_m(x)\gamma^{\mu_1\mu_2}\psi^m(y)\cdots, \bar\psi_m(x)\gamma^{\mu_1\cdots\mu_n}\psi^m(y)
\eeq
while in even dimensions, one must also account for chiral operators involving $\gamma^5$. Consequently, the nature of bulk fields in the holographic description necessarily involves higher-form fields for $d\geq 4$. For this reason, we will presently restrict our discussion to the case $d=3$, although much should have a natural generalization to arbitrary dimension. Note also that the 0-form source $A(x,y)$ in the fermionic model is parity-odd (i.e., a pseudoscalar), as opposed to the bosonic counterpart $\Scal(x,y)$, which was parity even. 

As was the case with the bosonic $U(N)$ model, the action above has a $CU(L_2)$ symmetry, under which $W_{\mu}$ transforms as a connection, while $A$ transforms as an adjoint scalar. We may split the connection 
\beq
W_{\mu} = W^{(0)}_{\mu}+\Cont_{\mu}
\eeq
into a flat piece $W^{(0)}$ and the tensorial piece $\Cont$. In the bosonic model, we used a field redefinition symmetry to absorb $\Cont$ into the 0-form $B$. However, in the fermionic case, we do not have the liberty to do so because of the gamma matrix structures involved. Therefore, we must keep both $A$ as well as $\Cont_{\mu}$ as the tensorial perturbations away from the free fixed point. The renormalization group extends the field theory sources into fields $\mathcal{A}$ and $\cConn$ living on the one-higher dimensional RG mapping space. The ERG equations take the form of bulk equations of motion, given by
\newcommand{\cA}{\mathcal{A}}
\beqn
\mathcal{F}^{(0)} &\equiv& \boldsymbol{d}\cConn^{(0)}+\cConn^{(0)}\wedge \cConn^{(0)} = 0\\
\left[\cD,\cA\right]_{\bl} &\equiv&  \boldsymbol{d}\cA + \left[\cConn,\cA\right]_{\bl} = \boldsymbol{\beta}^{(\cA)}\\
\mathcal{F} &\equiv& \boldsymbol{d}\cConn+\cConn\wedge \cConn = \boldsymbol{\beta}^{(\cConn)}
\eeqn
where the beta functions are given by
\beq
\boldsymbol{\beta}^{(\cA)}= \left(\cA\bl\Delta^{\mu}\bl\cCont_{\mu}+\cCont_{\mu}\bl\Delta^{\mu}\bl\cA+\epsilon^{\mu\nu\lambda}\cCont_{\mu}\bl\Delta_{\nu}\bl\cCont_{\lambda}\right)dz + \boldsymbol{\beta}^{(\cA)}_{\mu}dx^{\mu}
\eeq
\beqn
\boldsymbol{\beta}^{(\cConn)}&=& \left(\cA\bl\Delta_{\mu}\bl\cA+ \epsilon^{\mu\nu\lambda}(\cA\bl\Delta_{\nu}\bl\cCont_{\lambda}+\cCont_{\nu}\bl\Delta^{\lambda}\bl\cA)+\cCont_{\nu}\bl\Delta^{\nu}\bl\cCont_{\mu}-\cCont_{\nu}\bl\Delta_{\mu}\bl\cCont^{\nu}\right.\nonumber\\
&+&\left.\cCont_{\mu}\bl\Delta^{\nu}\bl\cCont_{\nu}\right)dz\wedge dx^{\mu} + \boldsymbol{\beta}^{(\cConn)}_{\mu\nu}dx^{\mu}\wedge dx^{\nu}
\eeqn
where the transverse components of the beta functions above are determined by the Bianchi identities. The interpretation of these equations as the Hamilton equations for radial evolution along $z$ goes through straightforwardly. Once again, the Hamiltonian turns out to be linear in momenta, and we may write a phase space action which reproduces all the RG equations as the corresponding equations of motion
\beq
I = \int_{\infty}^{\epsilon} dz\;\mathrm{Tr}\Big\{\mathcal{P}^I\left([\cD_I,\cA]_{\bl}-\boldsymbol{\beta}^{(\cA)}_I\right)+\mathcal{P}^{IJ}\left(\mathcal{F}_{IJ}-\boldsymbol{\beta}^{(\cConn)}_{IJ}\right) -iN\Delta^I\bl\cCont_I\Big\}
\eeq
To compute the boundary correlation functions, we need to evaluate this action on-shell
\beq
I_{o.s} =iN \int_{\epsilon}^{\infty} dz\;\mathrm{Tr}\;\Delta^{\mu}\cCont_{\mu}
\eeq
where we have made a convenient gauge choice $\cCont_z = 0$. To proceed, we need to obtain $\cCont_{\mu}(z)$ by solving the equations of motion (once again, focussing on the $z$-component)
\beqn
\left[\cD^{(0)}_z,\cA\right]_{\bl} &=&  \boldsymbol{\beta}^{(\cA)}_{z}\\
\left[\cD^{(0)}_z,\cCont_{\mu}\right]_{\bl} &=& \boldsymbol{\beta}^{(\cConn)}_{z\mu}
\eeqn
It is most convenient to solve these equations iteratively
\beqn
\cA &=& \cA^{(0)}+\cA^{(1)}+\cA^{(2)}+\cdots\\
\cCont_{\mu} &=& \cCont_{\mu}^{(0)}+\cCont_{\mu}^{(1)}+\cCont_{\mu}^{(2)}+\cdots
\eeqn
The background values are taken to be $\cA^{(0)} = \cCont_{\mu}^{(0)} = 0$, as these correspond to the free fixed point. We will take the boundary to lie at $z=\epsilon$, and fix the boundary values to be
\beq
\cA(\epsilon;x,y) = a^{(0)}(x,y),\;\;\cCont_{\mu}(\epsilon;x,y) = \hat{w}^{(0)}_{\mu}(x,y)
\eeq
As in the bosonic case, the other boundary condition is the interior boundary condition, i.e.
\beq
\lim_{z\to \infty} \mathcal{P}(z;x,y) = \lim_{z\to \infty} \mathcal{P}^{\mu}(z;x,y) = 0
\eeq
where $\mathcal{P}$ and $\mathcal{P}^{\mu}$ are the momenta corresponding to $\cA$ and $\cCont$. The first order sources are easily solved for 
\beq
\cA^{(1)} = K\bl a^{(0)}\bl K^{-1},\;\;\;\cCont_{\mu}^{(1)}= K\bl \widehat{w}_{\mu}^{(0)}\bl K^{-1}
\eeq
where $K$ is the boundary to bulk Wilson line defined in the main text.
The equations for the $k$th order sources take the form
\beq
\left[\cD^{(0)}_z,\cA^{(k)}\right]_{\bl} =  \Phi_{\cA}^{(k)}\label{Aeom}
\eeq
\beq
\left[\cD^{(0)}_z,\cCont_{\mu}^{(k)}\right]_{\bl} = \Phi_{\cConn}^{(k)} \label{Weom}
\eeq
where the inhomogenous terms $\Phi_{\cA}^{(k)}$ and $\Phi_{\cConn}^{(k)}$ are made up of lower order sources. In terms of the bulk-to-bulk Wilson line $G$ (defined in the main text), 
we find that equations \eqref{Aeom} and \eqref{Weom} can be solved as 
\beqn
\cA^{(k)}(z) &=& \int_{\epsilon}^{\infty}dz'\;G(z,z')\bl\Phi^{(k)}_{\cA}\bl G^{-1}(z',z)\nonumber\\
\cCont^{(k)}(z) &=& \int_{\epsilon}^{\infty}dz'\;G(z,z')\bl\Phi^{(k)}_{\cConn}\bl G^{-1}(z',z)
\eeqn
All that remains is to plug these solutions into the on-shell action and perform the $z$ integrations. This computation proceeds in exactly the same way as the bosonic case, and so we do not show all the details here. The on-shell action at the $k$th order can be massaged into the form
\beq
I_{o.s}^{(k)} = iN\int_{\epsilon}^{\infty}dz_1\int_{\epsilon}^{z_1}dz_2\cdots\int_{\epsilon}^{z_{k-1}}dz_k\;\mathrm{Tr}\left(\slashed{H}(z_1)\bl(a^{(0)}+\gamma^{\mu}\widehat{w}_{\mu}^{(0)})\cdots \slashed{H}(z_k)\bl(a^{(0)}+\gamma^{\sigma}\widehat{w}_{\sigma}^{(0)})+\mathrm{permutations}\right)\label{Diracaction1}
\eeq
where we have defined $\slashed{H}(z) = K^{-1}(z)\bl\gamma^{\mu}\Delta_{\mu}(z)\bl K(z)$. Defining the Wilsonian Green function 
\beq
g_{\mu}(z;x,y) = \int_{\epsilon}^z dz'\;K^{-1}(z')\bl \Delta_{\mu}(z')\bl K(z')
\eeq
\beq
g^{(0)}_{\mu}(x,y) = g_{\mu}(\infty;x,y)
\eeq
we find (after performing the $z$ integrations) that equation \eqref{Diracaction1} becomes
\beq
I_{o.s}^{(k)} = i\frac{N}{k}\mathrm{Tr}\left(\gamma^{\nu}g_{\nu}^{(0)}\bl(a^{(0)}+\gamma^{\mu}\widehat{w}_{\mu}^{(0)})\right)^k
\eeq
Thus, the bulk action evaluated on-shell reproduces the field theory generating functional (up to source-independent boundary terms)
\beq
I_{o.s.} = iN\mathrm{Tr\;ln}\;\left((\gamma^{\mu}g^{(0)}_{\mu})^{-1}-a_{(0)}-\gamma^{\mu}\widehat{w}^{(0)}_{\mu}\right)
\eeq


\providecommand{\href}[2]{#2}\begingroup\raggedright\endgroup

\end{document}